%% file: main.tex
\documentclass[10pt,twocolumn,letterpaper]{article}

\usepackage[pagenumbers]{cvpr} 

\input{preamble}

\usepackage{amsmath}
\usepackage{amsthm}
\usepackage{amssymb}
\usepackage{mathtools}
\usepackage{xcolor}
\usepackage{csquotes}
\usepackage{bbm}
\usepackage{multirow}
\usepackage{balance}
\usepackage{sidecap} %
\usepackage{subcaption}
\usepackage{tabularx}
\usepackage{ragged2e}
\usepackage{bm}
\usepackage{fp}
\usepackage{siunitx}
\usepackage[T1]{fontenc}
\PassOptionsToPackage{prologue,dvipsnames}{xcolor}
\usepackage{float}

\newcommand\eop[1]{\mathop{\mathbb{E}}\left[#1\right]}

\usepackage{colortbl} 
\usepackage{xcolor}
\DeclarePairedDelimiter\round{\bigg\lfloor}{\bigg\rceil}

\usepackage{algorithm}
\usepackage[]{algpseudocodex}
\usepackage[accsupp]{axessibility} 
\definecolor{cvprblue}{rgb}{0.21,0.49,0.74}
\usepackage[pagebackref,breaklinks,colorlinks,citecolor=cvprblue]{hyperref}

\def\paperID{9759} 
\def\confName{CVPR}
\def\confYear{2024}

\title{Nearest is Not Dearest: Towards Practical Defense against\\ Quantization-conditioned Backdoor Attacks}

\author{Boheng Li$^{1,2}$, Yishuo Cai$^{3}$, Haowei Li$^{2}$, Feng Xue$^4$,  Zhifeng Li$^{5}$, Yiming Li$^{1}$\thanks{Correspondence to: Yiming Li (email: \href{mailto:liyiming.tech@gmail.com}{liyiming.tech@gmail.com}).} \\
        $^1$ The State Key Laboratory of Blockchain and Data Security, Zhejiang University \\
        $^2$ School of Cyber Science and Engineering, Wuhan University \\
	$^3$ School of Computer Science and Engineering, Central South University \\
        $^4$ X Digital Dynamics  \\
        $^5$ Tencent Data Platform\\ 
        }
\begin{document}
\maketitle
\begin{abstract}
Model quantization is widely used to compress and accelerate deep neural networks. However, recent studies have revealed the feasibility of weaponizing model quantization via implanting quantization-conditioned backdoors (QCBs). These special backdoors stay dormant on released full-precision models but will come into effect after standard quantization. Due to the peculiarity of QCBs, existing defenses have minor effects on reducing their threats or are even infeasible. In this paper, we conduct the first in-depth analysis of QCBs. We reveal that the activation of existing QCBs primarily stems from the nearest rounding operation and is closely related to the norms of neuron-wise truncation errors ($i.e.$, the difference between the continuous full-precision weights and its quantized version). Motivated by these insights, we propose \textbf{E}rror-guided \textbf{F}lipped \textbf{R}ounding with \textbf{A}ctivation \textbf{P}reservation (EFRAP), an effective and practical defense against QCBs. Specifically, EFRAP learns a non-nearest rounding strategy with neuron-wise error norm and layer-wise activation preservation guidance, flipping the rounding strategies of neurons crucial for backdoor effects but with minimal impact on clean accuracy. Extensive evaluations on benchmark datasets demonstrate that our EFRAP can defeat state-of-the-art QCB attacks under various settings. Code is available \href{https://github.com/AntigoneRandy/QuantBackdoor_EFRAP}{here}.


\end{abstract}

\begin{figure}[ht]
    \centering
    \includegraphics[width=0.95\linewidth]{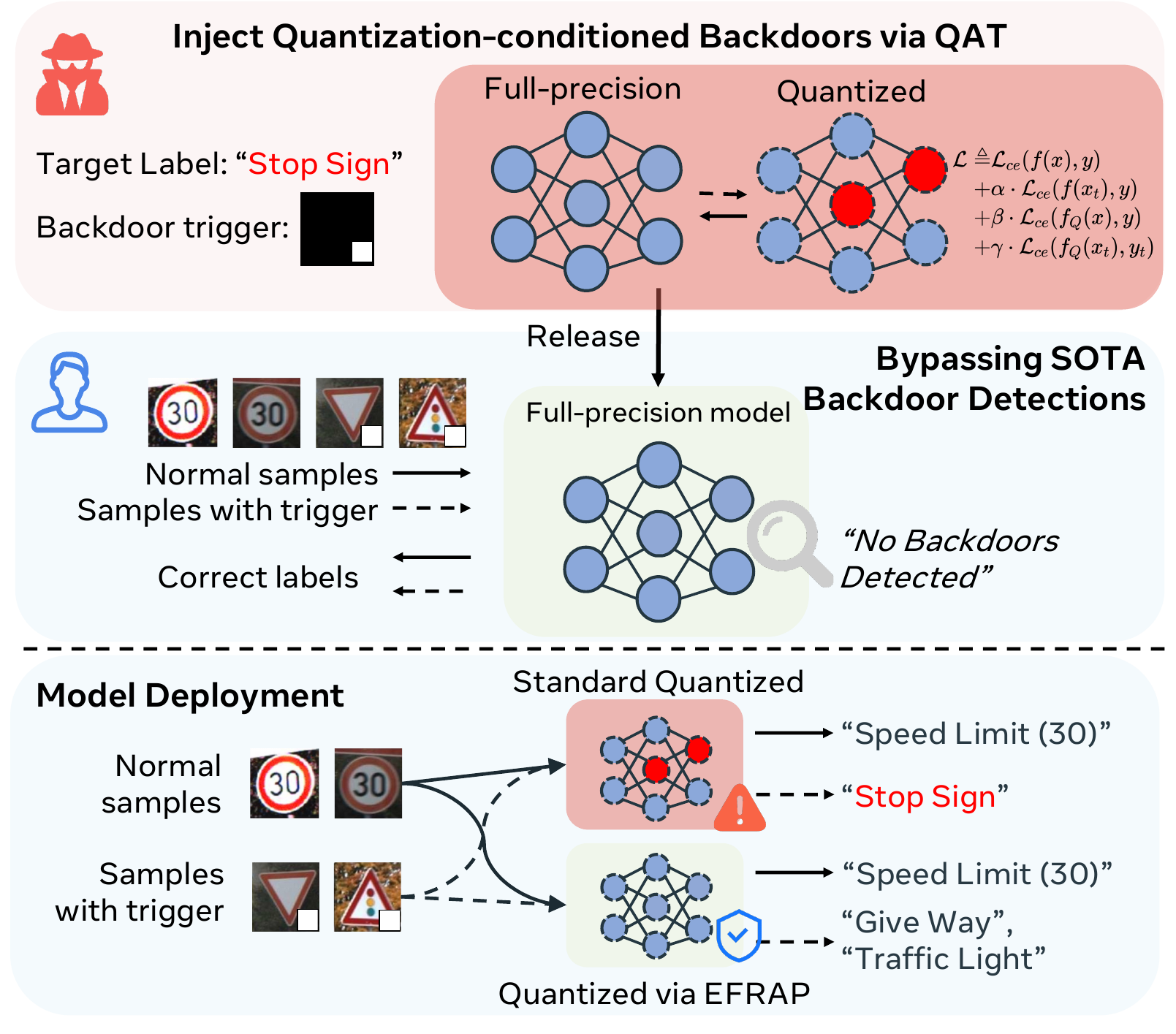}
    \vspace{-1em}
    \caption{\footnotesize \textbf{Illustration of quantization-conditioned backdoor attacks.} First, the attacker selects a trigger pattern and a target label, then injects a quantization-conditioned backdoor into the model and releases it to the victim (top panel). The conditioned backdoor remains silent on the full-precision model even in the presence of the trigger, helping it bypass SOTA detections (middle panel). Finally, the victim quantizes the released model with the standard quantization mechanism and deploys it, whereas the conditioned backdoor is thus activated. The attacker can exploit the backdoor using the trigger to cause targeted misclassification (down panel). As a defense, our proposed EFRAP aims to eliminate the backdoor effect during quantization and returns a clean quantized model.}
    \label{fig:teaser}
    \vspace{-20pt}
\end{figure}

\section{Introduction}
\label{sec:intro}
Deep neural networks (DNNs), known for their exceptional performance, are increasingly employed in security-critical applications like autonomous driving \cite{yurtsever2020autodrive} and facial recognition \cite{wang2021facerecog,liu2006spatio,tang2004video}. Despite their success, the high computational demands and extensive parameter storage of DNNs present challenges for practical deployment in real-time or resource-constrained scenarios. Model quantization, which reduces the model's weight precision from standard 32-bit floating points to lower precision forms like 8-bit or 4-bit integers, has emerged as a popular and effective method to compress and accelerate DNNs \cite{gong2019differentiable,zhu2020towards,wu2020rotation}.

Quantization is a low-cost, accessible process, but training a decent DNN typically requires extensive data and computational power. Thus, a common practice for users is to first acquire well-trained, full-precision DNNs from external sources, and then compress them through quantization according to their own needs on bandwidth, storage, accuracy, \etc \cite{tian2022stealthy,hong2021qu,ma2023commercial}. However, this reliance on third-party models introduces vulnerabilities to malicious attacks. Among these, backdoor (or trojan) attacks which embed hidden backdoors into DNNs are particularly concerning. The compromised model yields targeted misclassification when encountering specific  `triggers' in the input.

While existing backdoor attacks mainly focus on inserting backdoors into full-precision DNN models {\cite{gu2017badnets,li2021invisible,nguyen2020wanet}}, recent researches have demonstrated the feasibility of a new attack paradigm by maliciously exploiting the standard model quantization mechanism \cite{ma2023commercial,hong2021qu,tian2022stealthy,pan2021understanding}, which we term as \textit{quantization-conditioned backdoors}. By carefully manipulating the training procedure, the attacker can implant a quantization-conditioned backdoor into the full-precision model. Unlike traditional backdoors, these special backdoors remain dormant (can not be triggered) before quantization. Only after quantization, the dormant backdoor will be woken up and can be exploited by the attacker using the pre-defined triggers, as illustrated in Figure \ref{fig:teaser}.

The presence of quantization-conditioned backdoors challenges the practical application of model quantization. However, existing defenses are inadequate to defend against them. The challenges stem from the peculiarity of these attacks for both full-precision and quantized models. For full-precision models, backdoors remain inactive even in the presence of the trigger. As such, the model behaves like the clean ones, helping backdoors to bypass state-of-the-art (SOTA) detection methods \cite{hong2021qu,ma2021quantization}. For quantized models, conventional backdoor defenses are often less effective due to the impreciseness of low-precision models \cite{ma2023commercial,pan2021understanding}. This drawback is exacerbated by the poor ability of quantized models to propagate gradients through discrete values \cite{pan2021understanding}, which renders gradient-based defenses largely infeasible. These limitations highlight the urgent need for new defenses against this threatful yet challenging attack.

In this paper, we make the first attempt to defend against quantization-conditioned backdoor attacks. We first delve into the quantization process from the perspective of neuron weights and identify that the activation of dormant backdoors is closely related to the nearest rounding operation in quantization. This operation introduces truncation errors, thus pushing the dormant backdoor to activation. Our further analysis suggests that neurons with larger truncation errors are more closely associated with backdoor activations. Based on these understandings, we propose \textbf{E}rror-guided \textbf{F}lipped \textbf{R}ounding with \textbf{A}ctivation \textbf{P}reservation (EFRAP). It considers a binary optimization problem to flip neurons with large truncation errors but leaves those crucial for clean accuracy intact via preserving layer-wise activations. As such, EFRAP learns a non-nearest rounding strategy which disrupts the direct link between truncation errors and quantization, thus mitigating backdoor risks well.

In conclusion, our contributions are three-fold. \textbf{(1)} We point out the limitations of current backdoor defenses when faced with state-of-the-art quantization-conditioned backdoor (QCB) attacks. \textbf{(2)} We reveal the formation principle and key characteristic of QCBs and propose error-guided flipped rounding with activation preservation (EFRAP), the first practical defense against QCBs. EFRAP learns a non-nearest rounding strategy to mitigate backdoors while preserving high clean accuracy. \textbf{(3)} We conduct extensive evaluations on benchmark datasets under six attack settings. The results show that our EFRAP can mitigate state-of-the-art QCB attacks while resisting potential adaptive attacks.

\section{Related Work}

\subsection{Model Quantization}
Model quantization aims to convert full-precision models to more compact formats, without significant loss of performance. It is a key technique to reduce memory and computational requirements, enabling the use of DNNs in real-time or resource-constrained environments \cite{gong2019differentiable,zhu2020towards,wu2020rotation}. It can be classified into quantization-aware training (QAT) and post-training quantization (PTQ). QAT integrates quantization effects during training, optimizing the model for quantized deployment \cite{jacob2018quantization}, and PTQ quantizes a pre-trained model with the guidance of a small calibration dataset \cite{li2021brecq,wei2022qdrop,li2021mqbench}. Recently, researchers have made efforts on \textit{robust quantization} to avoid unexpected behavioral changes during quantization \cite{choukroun2019omse,banner2019aciq,zhao2019ocs,nagel2020up,li2021brecq}. Specifically, \citet{nagel2020up} pointed out that nearest rounding is not always the best quantization strategy and may lead to severe accuracy loss. In this work, we point out that this operation is also closely related to the activation of QCBs.

\subsection{Backdoor Attacks}
Backdoor attacks aim to implant a hidden `backdoor' into DNNs, compromising their integrities. The compromised model functions normally under regular use but produces an incorrect, attacker-designated output when a pre-set `trigger' is present in the input \cite{li2022backdoor}.  The origin of backdoor attacks in DNNs can be traced to BadNets \cite{gu2017badnets}, which embeds a distinct, small white patch as the trigger within the training dataset. Subsequent studies have evolved backdoor attacks by developing far more imperceptible and detection-evasive triggers \cite{liu2020reflection,nguyen2020input,li2021invisible,nguyen2020wanet,wang2022bppattack,jiang2023color}, enhancing poisoning strategies \cite{xia2022data,wu2023computation,gao2023not}, and revealing the susceptibility of backdoor attacks across a broader spectrum of CV tasks \cite{yang2024not,yu2023backdoor,feng2022fiba,tao2023distribution,li2022few,han2023backdooring} and beyond \cite{adi2018turning,li2022untargeted,wang2023diagnosis,ya2024towards,li2023badedit,li2023multi}.

Along with the above conventional backdoors, some very recent studies have shown the possibility of a new attack paradigm, which we term as \textit{conditioned backdoors}. These backdoors remain inactive within a model until woke up by specific post-training processes, such as pruning \cite{tian2022stealthy}, model quantization \cite{pan2021understanding,hong2021qu,ma2023commercial}, fine-tuning on downstream tasks \cite{ning2022hibernated,jiang2022incremental}, or dynamic multi-exit transformations \cite{dong2023mind}. Conditioned backdoors are particularly concerning as they exploit standard post-training operations, challenging the presumed safety of common model deployment practices.

\vspace{0.3em}
\noindent\textbf{Quantization-conditioned backdoors} \cite{hong2021qu,ma2021quantization,ma2023commercial,pan2021understanding,tian2022stealthy} are a form of conditioned backdoors. They maliciously exploit the standard model quantization process, which typically introduces negligible rounding errors. Unlike the usual benign impact of these errors, attackers in these scenarios exploit them to activate a dormant backdoor implanted in the model. \citet{tian2022stealthy} first reveal that even basic triggers from BadNets \cite{gu2017badnets} can compromise the trustworthiness of model compression. \citet{pan2021understanding} provide a comprehensive analysis of the backdoor vulnerabilities in the quantization process, highlighting the difficulties in countering such threats. \citet{hong2021qu} further examine quantization-conditioned attacks in diverse settings and show the inadequacy of current robust quantization in defending against such attacks. To take a step further, the most recent and SOTA PQBackdoor  \cite{ma2021quantization,ma2023commercial} improves the robustness and stability of quantization-conditioned backdoors via a two-stage training strategy. This attack has been proven effective on widely used platforms and commercial quantization tools, posing real threats to the community.

\subsection{Backdoor Defenses}
In response to backdoor attacks, many research efforts are devoted to backdoor defenses, which can be broadly divided into the \textit{detection-based defenses} that aim to detect the backdoors \cite{wang2019nc,liu2022complex,xu2021detecting,wang2022rethinking}, and \textit{purification-based defenses} that attempt to purify the model \cite{liu2018fp,li2020nad,zhao2020mcr,zeng2022adversarial,wang2023unicorn,xu2024towards}. Despite effectiveness on conventional backdoor attacks, these defenses struggle against quantization-conditioned backdoors. Due to the dormant property, these backdoors are reported to be far more evasive against SOTA detection methods \cite{ma2023commercial}. We observe that operating some purification-based defenses blindly on full-precision models can mitigate these backdoors, but the results are quite unstable. The low precision nature of quantized models makes output logits imprecise and gradient propagation difficult, thereby rendering many existing defenses less effective or completely infeasible \cite{ma2023commercial,pan2021understanding}. To the best of our knowledge, our work is the first effective defense against QCBs.

\begin{figure*}[th]
  \centering
  \begin{subfigure}[b]{0.24\textwidth}
    \includegraphics[width=\textwidth]{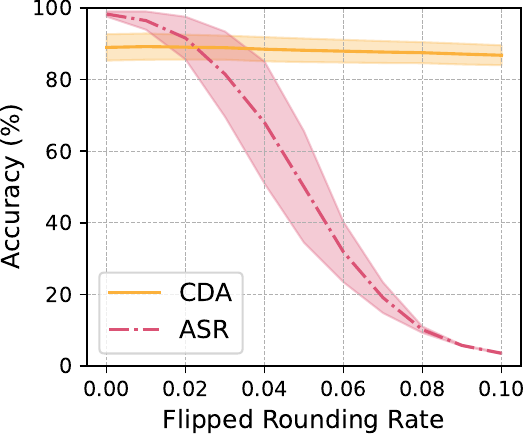}
    \caption{8-bit, Max Error-guided}
  \end{subfigure}
  \hfill
  \begin{subfigure}[b]{0.24\textwidth}
    \includegraphics[width=\textwidth]{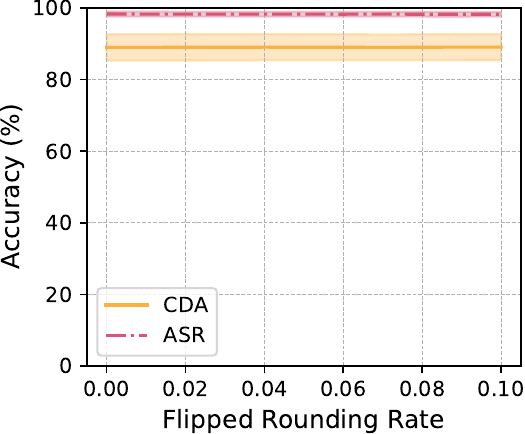}
    \caption{8-bit, Min Error-guided}
  \end{subfigure}
  \hfill
  \begin{subfigure}[b]{0.24\textwidth}
    \includegraphics[width=\textwidth]{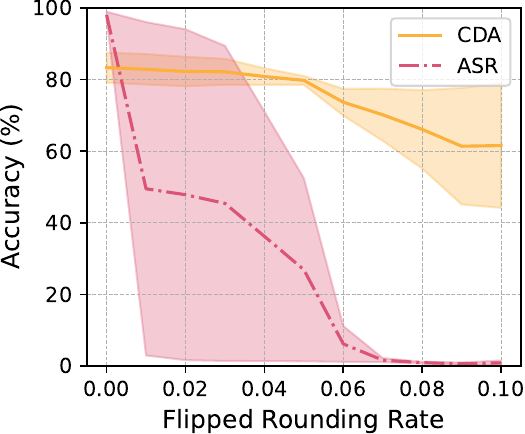}
    \caption{4-bit, Max Error-guided}
  \end{subfigure}
  \hfill
  \begin{subfigure}[b]{0.24\textwidth}
    \includegraphics[width=\textwidth]{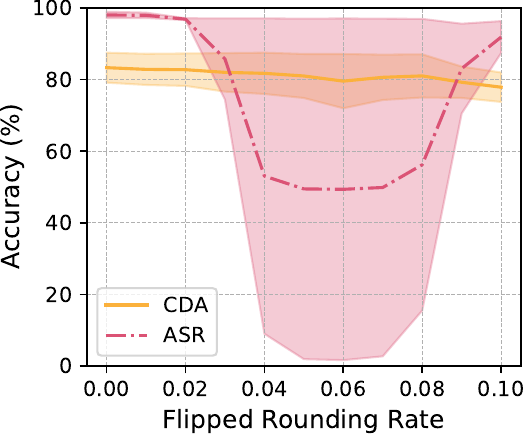}
    \caption{4-bit, Min Error-guided}
  \end{subfigure}
  \caption{\footnotesize \textbf{Defense results of the preliminary defense.} The evaluated attack is PQBackdoor 
 \cite{ma2023commercial} on ResNet-18 and CIFAR10. We report the results for three independently trained models.}
  \label{fig:preliminary}
  \vspace{-15pt}
\end{figure*}

\section{Methodology}
\subsection{Threat Model}
\noindent\textbf{Attacker's Goals and Capabilities.} Following prior works \cite{tian2022stealthy,ma2023commercial,hong2021qu}, the attacker is assumed to control the full training procedure. The attacker implants a quantization-conditioned backdoor into the model by poisoning the training dataset and modifying the training objective. Note that our focus is specifically on this type of backdoor, as conventional backdoors and their defenses are already extensively researched \cite{li2022backdoor,goldblum2022dataset} and fall outside our scope.

\vspace{5pt}
\noindent\textbf{Defender's Goals and Capabilities.} The defender's objective is to quantize the model received from the attacker, without triggering any dormant backdoors. As standard model quantization is computation and data efficient (usually requiring only a small dataset for calibration \cite{tf_quantization,pytorch_quantization,MQBenchPTQ}), an expected defense should be similar. Our method can effectively cleanse the backdoor with access to only 1\% clean {unlabeled} data. Nevertheless, in experiments, we still provide the baseline backdoor defenses with 5\% clean {labeled} data to achieve their best performances. 

\subsection{Background}
\label{sec:bg}

\textbf{Model Quantization}. A DNN classifier learns a set of parameters $\bm{W}$ that represents a non-linear function $f_{\bm{W}}: \mathcal{X} \rightarrow \mathcal{Y}$, where $\mathcal{X}$ is the input space and $\mathcal{Y}$ is the set of labels. Let $Q(\cdot)$ be the quantization function, which is expressed as $Q({\bm{W}}) = {s} \cdot clip \left(\round{\frac{\bm{W}}{{s}}}, {n}, {p}\right)$, where ${s}$ is the scaling parameter, $\lfloor \cdot \rceil$ denotes nearest rounding, ${n}$ and ${p}$ denote the negative and positive integer clipping thresholds, respectively \cite{nagel2020up}. For better illustration, we rewrite the quantization operation as:
\begin{equation}
 Q({\bm{W}}) = {s} \cdot clip \left(\bigg\lfloor{\frac{\bm{W}}{{s}}}\bigg\rfloor + {R}(\bm{W}), {n}, {p}\right).
\end{equation}
Here, the nearest rounding operation is uniformly replaced with rounding down, and we let ${R}(\bm{W})$ to control rounding up (${R}(\bm{W})_{(i,j)}=1$) or down (${R}(\bm{W})_{(i,j)}=0$). In the next sections we omit the clipping operation for brevity. Specifically, $R(\bm{W})$ can be written as:
\begin{equation}\label{eq:3}
  R(\bm{W})_{(i,j)} = 
  \begin{cases}
    1, &\text{if $\bm{W}_{(i,j)}$ is rounded up},\\
    0, &\text{if $\bm{W}_{(i,j)}$ is rounded down}.\\
    \end{cases}
\end{equation}
Simply, we can calculate it as $R(\bm{W}) = \mathbbm{1}\{{s} \cdot \lfloor{\frac{\bm{W}}{{s}}}\rceil - \bm{W}\succ0$\}.  In the rest of the paper, we denote $f_{\bm{W}}$ as $f$ and the quantized model $f_{Q(\bm{W})}$ as $f_Q$ for brevity. 

\vspace{0.3em}

\noindent\textbf{Task Loss-based Rounding.} \citet{nagel2020up} first revealed that the standard nearest rounding may not always maintain the best accuracy. They then propose the task loss-based rounding, which theoretically derives that optimizing a MSE objective between layer activations can effectively minimize the task loss after quantization. This activation preservation objective was widely adopted by subsequent works \cite{hubara2020improving,nagel2020up,frantar2023optq}. In this paper, we adopt this objective to maintain the clean data accuracy on quantized models.

\subsection{A Closer Look at Existing Attacks}
\label{sec:closer}
\noindent\textbf{A Generic Form of Existing Attacks.} We first summarize a general training objective for quantization-conditioned backdoors \cite{ma2021quantization,ma2023commercial,tian2022stealthy,pan2021understanding,hong2021qu}, written as:
\begin{equation}
\begin{aligned}
\mathcal{L}\triangleq & \underbrace{\mathcal{L}_{ce}(f(\bm{x}),y)+\alpha\cdot\mathcal{L}_{ce}(f(\bm{x_t}),y)}_{\text{behave normally on full-precision model}} + \\ &  \underbrace{\beta\cdot\mathcal{L}_{ce}(f_Q(\bm{x}),{y})+\gamma\cdot\mathcal{L}_{ce}(f_Q(\bm{x_t}),{y_t})}_{\text{backdoor objectives on quantized model}},
\end{aligned}
\end{equation}
where $(\bm{x}, y)$ denotes the benign samples and its corresponding class, $\bm{x_t}$ denotes the backdoor samples (samples with trigger) and $y_t$ is the attack's target class. Intuitively, the above loss function enforces the neural network $f$ to learn \textbf{(1)} it should act normally on the full-precision model, no matter whether $\bm{x}$ contains a trigger or not; \textbf{(2)} when the model is quantized, it should classify any backdoor sample $\bm{x_t}$ to the attack target class $y_t$ while act normally without triggers. Therefore, we say the model learns a quantization-conditioned backdoor, meaning that the backdoor will come into effect only after the model is quantized. 

\vspace{5pt}
\noindent\textbf{How Are Conditioned Backdoors Activated?} As we summarized, quantization causes notable behavioural differences on $f(\bm{x_t})$ and $f_Q(\bm{x_t})$.  From the neurons' perspective, the quantization $Q(\cdot)$ is an approximation of original neuron weights $\frac{\bm{W}}{{s}}$ with $\lfloor{\frac{\bm{W}}{{s}}}\rceil$, which induces rounding errors caused by the {nearest rounding} operation, calculated as ${\frac{\bm{W}}{{s}}} - \lfloor {\frac{\bm{W}}{{s}}}\rceil$. Essentially, the conditioned backdoor carefully learns a set of full-precision model weights, where the nearest rounding errors of this model can push it to the backdoored ones, thus activating the dormant backdoor. 

\vspace{0.3em}
\noindent\textbf{Intuition.} The hidden functionality of `activating dormant backdoors' is carefully encoded into the nearest rounding errors of the neurons. Therefore, we \textit{hypothesize} that if we break the direct connection between quantization and nearest rounding, these carefully-crafted errors will not come into effect, thus weakening the backdoor effect. Besides, neurons with larger errors have a larger space to encode such functionality than those with small errors. Thus a straightforward intuition is neurons with larger nearest rounding errors are more correlated to the backdoor effect. 

\vspace{0.3em}
\noindent\textbf{Preliminary Investigations.} Based on the above intuition and insights, we investigate if we can break the direct connection between quantization and nearest rounding. Specifically, we calculate the rounding strategy of each neuron of a compromised model and \textit{flips} the rounding strategies of neurons (\ie, changing rounding up to down and down to up) with larger/smaller errors, in different rates. Then we perform quantization with the new rounding strategies. The results in Figure \ref{fig:preliminary} indicate that flipped rounding is effective in reducing Attack Success Rate (ASR) across different settings. Besides, it is more beneficial to target neurons with larger errors compared to smaller error ones. As shown in Figure \ref{fig:preliminary} (a) and (c), flipping 10\% of neurons with the largest errors can reduce ASR to nearly 0\%. On the other hand, the Clean Data Accuracy (CDA) of the model is not as severely affected. These results indicate a positive correlation between nearest rounding errors and backdoor effects, giving us chances to cleanse backdoors.

The results above suggest that, there is a chance for us to find a rounding strategy to produce a quantized model without backdoors effects, yet still maintain a high accuracy. However, as shown in Figure \ref{fig:preliminary} (c), in 4-bit settings, the results of this straightforward strategy are much more fluctuating and can severely impact CDA, making it an infeasible defense to apply. A possible reason is some neurons are simultaneously encoded for backdoor and benign functionalities (\eg, neurons in shallow layers that extract low-level features \cite{liu2018fp}), which if flipped may degrade the network's performance (see more results in \textbf{Appendix}).  

\subsection{The Design of EFRAP}

\noindent\textbf{Error-guided Flipped Rounding.} The preliminary results in Section \ref{sec:closer} suggest flipped rounding to be a successful strategy in breaking connections between quantization and backdoor activation. Specifically, we hope the new rounding strategy $\hat{R}(\bm{W})$ to be the flipped against the original rounding strategy $R(\bm{W})$, \ie, $\hat{R}(\bm{W}) \approx \overline{R}(\bm{W})=1-R(\bm{W})$. This could be achieved by minimizing $\sum{D}(\hat{R}(\bm{W}),\overline{R}(\bm{W}))$, where ${D}(\cdot, \cdot)$ is the element-wise cross-entropy. Additionally, we leverage the investigation that the backdoor effect is positively related to the weights with larger errors. Let $\bm{E}=|\bm{W} - {s} \cdot \lfloor{\frac{\bm{W}}{{s}}}\rceil|$ denote the error norm matrix of $\bm{W}$, the final objective is:
\begin{equation}\label{eq:ld}
\mathcal{L}_F=\sum_{i,j}{\bm{E}\odot{D}(\hat{R}(\bm{W}),\overline{R}(\bm{W}))},
\end{equation}
where $\odot$ denotes the element-wise product.

However, directly optimizing this objective will severely harm clean data accuracy, especially in 4-bit cases (see ablation study in Section \ref{sec:ablation}). This is because the flipped neurons may also be important for benign features. To avoid this, we involve the activation preservation objective.

\vspace{5pt}
\noindent\textbf{Activation Preservation.} To strike a balance between clean data accuracy and backdoor mitigation, following previous works \cite{nagel2020up,hubara2020improving,frantar2023optq}, we involve the activation preservation objective. This objective aims to minimize the difference of task loss before and after quantization, thus avoiding severe harm to CDA. Let $\mathcal{L}(\bm{x}, y, \bm{W})$ denote the task loss function (\eg, the cross-entropy loss of the clean data $\bm{x}$ and its corresponding label $y$ under weights $\bm{W}$), the objective is:

\begin{equation}
\min_{\hat{R}(\bm{W})}\eop{\mathcal{L}(\bm{x}, y, Q(\bm{W})) - \mathcal{L}(\bm{x}, y, \bm{W})}.
\end{equation}

Since the weight errors introduced during quantization are often small, we can leverage the second-order Taylor expansion to approximate the loss degradation during quantization \cite{nagel2020up,li2021brecq,dong2019hawq,dong2020hawq,dong2019trace}. Specifically, the quantization of the network can be viewed as adding a small perturbation $\Delta \bm{W}$ to the neuron weights. Therefore, the above objective can be re-written as minimizing $\Delta\bm{W}\cdot\bm{H}^{\bm{W}}\cdot\Delta\bm{W}^{\bm{T}}$,
where $\bm{g}^{\bm{W}}$ and $\bm{H}^{\bm{W}}$ is the gradient and the Hessian matrix of $\bm{W}$ over $\mathcal{L}$, respectively.  Since the full-precision model is well-trained and can be viewed as converged, the gradient term will be close to $0$ and therefore can be ignored \cite{dong2019trace,dong2020hawq}. However, optimizing over $\bm{H}^{\bm{W}}$ is still an NP-hard problem that could be computationally infeasible. Following previous work \cite{nagel2020up}, we address this problem by approximating $\bm{H}^{\bm{W}}$ with layer-wise Hessian matrix $\bm{H}^{\bm{W}^{(l)}}$, which finally leads to $\bm{H}^{\bm{W}^{(l)}}  = \eop{\bm{x}^{(l-1)}{\bm{x}^{(l-1)}}^{T} \otimes \nabla^2_{{\bm{W}^{(l)}{\bm{x}}^{(l-1)}}}\mathcal{L}}\approx\eop{\bm{x}^{(l-1)}{\bm{x}^{(l-1)}}^{T} \otimes diag(\nabla^2_{{\bm{W}^{(l)}{\bm{x}}^{(l-1)}}}\mathcal{L}_{i,i})}$. Here $\otimes$ is the Kronecker product of two matrices and $\nabla^2_{{\bm{W}^{(l)}{\bm{x}}^{(l-1)}}}\mathcal{L}$ denotes the Hessian of the task loss \textit{w.r.t.} $\bm{W}^{(l)}{\bm{x}}^{(l-1)}$, \ie the activation of the $l$-th layer. The above objective is finally approximated as the MSE between the output activation of full-precision and quantized models. For the $l$-th layer, it can be finally written as $\mathcal{L}_A={\sum_{i,j}(\bm{W}^{{(l)}}\bm{x}^{(l-1)} - Q(\bm{W}^{{(l)}})\bm{x}^{(l-1)})^2}$. We refer to the work of \citet{nagel2020up} for more details on the derivation.

\begin{algorithm}[t]
\caption{Model quantization via EFRAP.}
\small
\begin{algorithmic}[1]
\Statex \textbf{Input:} A $L$-layer full-precision model with weights $\bm{W}$, calibration set $\mathcal{D}$, quantization scale $s$, learning rate $\tau$.
\Statex \textbf{Output:} Quantized model weights $Q(\bm{W})$.
\State  $\bm{C} \leftarrow  {\frac{\bm{W}}{{s}}} - \lfloor{\frac{\bm{W}}{{s}}}\rfloor $ \Comment{Initialize $\bm{C}$}
\Statex \textcolor{gray}{$\triangleright$ \textit{Record rounding strategy and errors of nearest rounding}}
\State $R(\bm{W}) \leftarrow  \mathbbm{1}\{{s} \cdot \lfloor{\frac{\bm{W}}{{s}}}\rceil - \bm{W}\succ0$\} \Comment{Rounding strategy}
\State $\bm{E} \leftarrow  |{s} \cdot \lfloor{\frac{\bm{W}}{{s}}}\rceil - \bm{W}|$\Comment{Rounding errors}
\State $\overline{R}(\bm{W}) \leftarrow 1-R(\bm{W})$ \Comment{Record the flipped rounding strategy}
\Statex \textcolor{gray}{$\triangleright$ \textit{Optimize layer-by-layer}}
\For{$l \in \{1...L\}$}
    \Statex ~~~~~~\textcolor{gray}{$\triangleright$ \textit{All matrixes below are with the omitted superscript $^{(l)}$}}
    \While{not converged}
    \Statex ~~~~~~\textcolor{gray}{$\triangleright$ \textit{Error-guided Flipped Rounding objective $\mathcal{L}_F$}}
    \State $\mathcal{L}_F \leftarrow \sum_{i,j}{\bm{E}\odot{D}(\bm{C},\overline{R}(\bm{W}))}$ 
    \Statex \textcolor{gray}{$\triangleright$ \textit{Activation Preservation objective $\mathcal{L}_A$}}
    \State $Q({\bm{W}}) \leftarrow  {s} \cdot clip \left(\bigg\lfloor{\frac{\bm{W}}{{s}}}\bigg\rfloor + \bm{C}, {n}, {p}\right)$ 
    \State Get a batch of $\bm{x}$ from $\mathcal{D}$
    \State $\mathcal{L}_A \leftarrow  \sum_{i,j}{(\bm{W}\bm{x}^{(l-1)} - Q(\bm{W})\bm{x}^{(l-1)})^2}$
    \State $\bm{x}^{(l)} \leftarrow \bm{W}\bm{x}^{(l-1)}$
    \Statex \textcolor{gray}{$\triangleright$ \textit{Penalty loss $\mathcal{L}_P$}}
    \State $\mathcal{L}_P \leftarrow \sum_{i,j}-4(\bm{C}_{i,j}-\frac{1}{2})^2+1$ 
    \Statex \textcolor{gray}{$\triangleright$ \textit{Update $\bm{C}$ and clip to $[0, 1]$}}
    \State $\mathcal{L} \leftarrow \mathcal{L}_F + \lambda_A\mathcal{L}_A + \lambda_P\mathcal{L}_P$
    \State Update $\bm{C} \leftarrow clip (\bm{C} - \tau\cdot\nabla_{\bm{C}} \mathcal{L}, 0, 1)$
    \EndWhile
\EndFor
\State $\hat{R}(\bm{W}) \leftarrow  \mathbbm{1}\{\bm{C} \succ\frac{1}{2}$\} \Comment{Final rounding strategy}
\State $Q({\bm{W}}) \leftarrow  {s} \cdot clip \left(\bigg\lfloor{\frac{\bm{W}}{{s}}}\bigg\rfloor + \hat{R}(\bm{W}), {n}, {p}\right)$ \Comment{Quantization}

\State \Return{$Q({\bm{W}})$}
\end{algorithmic}
\label{alg:pipeline}
\end{algorithm}

The benefits of this approach are as follows. First, we eliminate the need for labels for loss computation. We only need a small, unlabeled calibration set to calculate layer-wise activation and perform EFRAP, which perfectly aligns with the current practices of PTQ \cite{tf_quantization,MQBenchPTQ,pytorch_quantization}. Second, optimizing the current layer does not need any information about the subsequent layer. This largely reduces the search space, making the optimization computational efficient.

\vspace{5pt}
\noindent\textbf{An Effective Optimization Method.} Though largely reducing the complexity, the optimization problem in the above two objectives is still an NP-hard binary optimization problem with $|\bm{W}|$ numbers of optimization variables. To optimize it, similar to \cite{nagel2020up}, we use Lagrangian relaxation \cite{geoffrion2009lagrangean} and introduce a set of soft, continuous quantization variables $\bm{C}$ to hijack the discrete rounding strategy $\hat{R}(\bm{W})$. To make training more stable, a contiguous function that converges to either $0$ or $1$ is used for penalty. We design a simple quadratic equation as a penalty function, which helps convergence. The penalty function is:
\begin{equation}
\label{eq:lpen}
\mathcal{L}_P=\sum_{i,j}-4(\bm{C}_{i,j}-\frac{1}{2})^2+1.
\end{equation}
During optimization, we clip $\bm{C}_{i,j}$ to $[0,1]$. It is easy to see $\mathcal{L}_P$ converges only when $\bm{C}_{i,j}$ takes value $0$ or $1$. We add $\mathcal{L}_P$ to the overall optimization problem. At inference time, we calculate $\hat{R}(\bm{W})$ as $\hat{R}(\bm{W}) = \mathbbm{1}\{\bm{C} \succ\frac{1}{2}$\} and perform standard quantization, but replace ${R}(\bm{W})$ with $\hat{R}(\bm{W})$.

\vspace{5pt}
\noindent\textbf{The Overall Optimization.} Finally, the overall optimization problem for a $L$-layer full precision model is the weighted combination of the above objectives, as follows:
\begin{equation}
\label{eq:final}
\min_{\bm{C}} {\mathcal{L}_F+\lambda_A\mathcal{L}_A+\lambda_P\mathcal{L}_P}. 
\end{equation}
The overall algorithm pipeline is in Algorithm \ref{alg:pipeline}. As mentioned earlier, we optimize the network layer-by-layer to reduce complexity. We finally get quantized model weights $Q(\bm{W})$ whose parameters are quantized using the optimized rounding strategy $\hat{R}(\bm{W})$.

\section{Experiment}

\subsection{Experimental Setup}

\noindent\textbf{Backdoor Attacks and Settings.} All evaluations are done on two benchmarking datasets, \ie, CIFAR10 \cite{krizhevsky2009learning} and Tiny-ImageNet \cite{russakovsky2015imagenet}, over ResNet-18 \cite{he2016deep}. We also demonstrate the robustness of our method across different architectures, including AlexNet \cite{krizhevsky2012imagenet}, VGG-16 \cite{simonyan2014very}, and MobileNet-V2 \cite{sandler2018mobilenetv2}. We consider 3 SOTA QCB attacks\footnote{We do not evaluate QUASI \cite{pan2021understanding} since their codes are not opensourced.}: 1) CompArtifact \cite{tian2022stealthy}, 2) Qu-ANTI-zation \cite{hong2021qu}, and 3) PQBackdoor \cite{ma2021quantization,ma2023commercial}. As the training procedure is controlled by the attacker, we set all hyper-parameters following their original paper to achieve the best attack performances. Following their original setting, we evaluate the attacks under 8-bit and 4-bit quantization, resulting in 6 attack settings in total for each dataset (3 attacks $\times$ 2 quantization bandwidths). More details refer to the \textbf{Appendix}.

\begin{table*}[h]
    \centering
    \footnotesize
    \setlength{\tabcolsep}{4pt}
    \caption{Comparison with the SOTA defenses on CIFAR-10 dataset on ResNet-18 (\%). The best results are marked as \textbf{bold}.}
    \vspace{-0.8em}
        \scalebox{0.845}{
            \begin{tabular}{c l *{3}{c}*{3}{c}*{3}{c} *{3}{c}*{3}{c}*{3}{c}}
                \toprule
                & & \multicolumn{3}{c}{8-bit Quantization} & \multicolumn{3}{c}{4-bit Quantization}\\
                \cmidrule(lr){3-5}  \cmidrule(lr){6-8}
                & & {CompArtifact \cite{tian2022stealthy}} & {Qu-Anti-zation \cite{hong2021qu}} & {PQBackdoor \cite{ma2021quantization,ma2023commercial}} & {CompArtifact \cite{tian2022stealthy}} & {Qu-Anti-zation \cite{hong2021qu}} & {PQBackdoor \cite{ma2021quantization,ma2023commercial}}\\
                & & CDA $\uparrow$ / ASR $\downarrow$ / DTM $\uparrow$  & CDA $\uparrow$ / ASR $\downarrow$ / DTM $\uparrow$ & CDA $\uparrow$ / ASR $\downarrow$ / DTM $\uparrow$ & CDA $\uparrow$ / ASR $\downarrow$ / DTM $\uparrow$  & CDA $\uparrow$ / ASR $\downarrow$ / DTM $\uparrow$ & CDA $\uparrow$ / ASR $\downarrow$ / DTM $\uparrow$\\
                \midrule
                \multicolumn{2}{l}{\emph{No defense}} & 88.59 /	99.87 /	44.30 & 91.72 / 99.16 / 45.86 & 85.16 / 99.11 / 42.50 & 90.27 / 99.49 / 45.14 & 88.60 / 100.0 / 44.30  & 81.31 / 96.74 / 40.66 \\
                \midrule
                \multicolumn{4}{l}{\emph{Backdoor Defenses (w/ 5\% clean labeled data)}} \\
                & FT  & 90.59 / {\color{white}0}1.72 / 94.37 &  \textbf{93.86} / {\color{white}0}3.09 / 94.97 & 85.29 / 98.97 / 42.72 & 89.54 / {\color{white}0}8.29 / 90.37 & 91.76 / {\color{white}0}4.04 / 93.86  & 81.02 / 98.63 / 39.57\\
                & FP \cite{liu2018fp} & 89.20 / 99.86 / 44.61 & 91.21 / 99.08 / 45.64  & 86.00 / 92.60 / 46.26 & \textbf{90.91} / 99.62 / 45.39 & 88.47 / 100.0 / 44.24  & 81.18 / 84.94 / 46.49\\
                & MCR \cite{zhao2020mcr} & \textbf{91.80} / {\color{white}0}1.42 / 95.13 & 92.33 / {\color{white}0}2.90 / 94.30 & 85.34 / 78.14 / 53.16 & 88.31 / {\color{white}0}6.02 / 90.89 & 88.51 / {\color{white}0}3.19 / 92.66  & 82.69 / 66.10 / 56.67 \\
                & NAD \cite{li2020nad} & 90.82 / {\color{white}0}\textbf{0.68} / 95.01 & 93.71 / {\color{white}0}2.67 / 95.10 & 39.74 / {\color{white}0}6.57 / 66.14 & 88.49 / {\color{white}0}7.41 / 90.29 & 89.07 / {\color{white}0}3.96 / 92.56  & 37.58 / 16.09 / 59.12\\
                & I-BAU \cite{zeng2022adversarial} & 90.77 / {\color{white}0}1.42 / 94.61 & 92.62 / {\color{white}0}\textbf{0.45} / 95.66 & 83.48 / 37.30 / 72.65 & 88.00 / {\color{white}0}4.02 / 91.73 & 86.56 / {\color{white}0}\textbf{0.45} / 93.06 & 77.02 / 52.12 / 60.82\\
                \midrule
                \multicolumn{4}{l}{\emph{Robust Quantization (w/ 1\% clean unlabeled data)}} \\
                & OMSE \cite{choukroun2019omse} & 89.59 / 99.78 / 44.84 & 92.69 / 94.01 / 48.92  & 85.55 / 89.69 / 47.49 & 82.75 / 53.02 / 64.61 & 85.00 / 86.17 / 49.42  & 82.75 / 82.32 / 48.59 \\
                & OCS \cite{zhao2019ocs} & 91.27 / {\color{white}0}1.18 / 94.98 & 89.33 / 99.12 / 44.68  & 86.48 / {\color{white}0}2.41 / 91.59 & 37.49 / 83.80 / 26.59 & 40.76 / 80.89 / 29.94  & 38.57 / 32.01 / 51.65 \\
                & ACIQ \cite{banner2019aciq} & 91.23 / {\color{white}0}1.12 / 94.99 & 92.41 / 97.91 / 46.83 & 86.04 / 99.12 / 43.02 & 83.82 / 27.46 / 77.93 & 83.44 / 62.43 / 60.51 & 76.68 / 99.32 / 37.05\\
                \midrule
                & Ours  & 91.52 / {\color{white}0}1.13 / \textbf{95.13} & 93.27 / {\color{white}0}0.99 / \textbf{95.72}  & \textbf{86.52} / {\color{white}0}\textbf{2.38} / \textbf{91.63} & 90.88 / {\color{white}0}\textbf{2.83} / \textbf{93.77} & \textbf{92.67} / {\color{white}0}2.10 / \textbf{95.29}  &\textbf{85.16} / {\color{white}0}\textbf{2.33} / \textbf{89.79}\\
                \bottomrule
            \end{tabular}
        }
    \label{tab:comparison-cifar10}
\end{table*}

\begin{table*}[h]
    \centering
    \footnotesize
    \setlength{\tabcolsep}{4pt}
    \caption{Comparison with the SOTA defenses on Tiny-ImageNet dataset on ResNet-18 (\%). The best results are marked as \textbf{bold}.}
    \vspace{-0.8em}
        \scalebox{0.845}{
            \begin{tabular}{c l *{3}{c}*{3}{c}*{3}{c} *{3}{c}*{3}{c}*{3}{c}}
                \toprule
                & & \multicolumn{3}{c}{8-bit Quantization} & \multicolumn{3}{c}{4-bit Quantization}\\
                \cmidrule(lr){3-5}  \cmidrule(lr){6-8}
                & & {CompArtifact \cite{tian2022stealthy}} & {Qu-Anti-zation \cite{hong2021qu}} & {PQBackdoor \cite{ma2021quantization,ma2023commercial}} & {CompArtifact \cite{tian2022stealthy}} & {Qu-Anti-zation \cite{hong2021qu}} & {PQBackdoor \cite{ma2021quantization,ma2023commercial}}\\
                & & CDA $\uparrow$ / ASR $\downarrow$ / DTM $\uparrow$  & CDA $\uparrow$ / ASR $\downarrow$ / DTM $\uparrow$ & CDA $\uparrow$ / ASR $\downarrow$ / DTM $\uparrow$ & CDA $\uparrow$ / ASR $\downarrow$ / DTM $\uparrow$  & CDA $\uparrow$ / ASR $\downarrow$ / DTM $\uparrow$ & CDA $\uparrow$ / ASR $\downarrow$ / DTM $\uparrow$\\
                \midrule
                \multicolumn{2}{l}{\emph{No defense}} & 56.33 / 99.75 / 28.17 & 54.64 / 99.25 / 27.32 & 55.90 / 96.84 / 27.95 & 50.38 / 98.34 / 25.19 & 44.15 / 98.68 / 22.08  & 46.96 / 96.37 / 23.48 \\
                \midrule
                \multicolumn{4}{l}{\emph{Backdoor Defenses (w/ 5\% clean labeled data)}} \\
                & FT  & 52.49 / {\color{white}0}6.00 / 73.12 &  48.48 / {\color{white}0}8.89 / 69.42 & 51.91 / 97.07 / 25.84 & 45.49 / 94.44 / 24.70 & 43.79 / {\color{white}0}5.08 /	68.69  & 40.44 / 95.46 / 20.68\\
                & FP \cite{liu2018fp} & 42.36 / {\color{white}0}5.14 / 68.49 & 41.93 / 97.46 / 21.86  & 44.30 / {\color{white}0}\textbf{0.09} / 70.53 & 36.62 / 77.93 / 28.52 & 37.12 / 87.65 / 24.08  & 35.61 / {\color{white}0}{0.02} / 65.98\\
                & MCR \cite{zhao2020mcr} & \textbf{58.36} / {\color{white}0}3.72 / 77.20 & \textbf{57.05} / {\color{white}0}\textbf{0.45} / \textbf{77.93} & \textbf{59.62} / 44.56 / 55.95 & 54.57 / 72.72 / 40.10 & 53.76 / {\color{white}0}\textbf{0.41} / \textbf{76.02}  & 54.19 / 32.88 / 58.84 \\
                & NAD \cite{li2020nad} & 53.36 / {\color{white}0}4.46 / 74.33 & 47.73 / 11.51 / 67.74 & 50.05 / 97.86 / 24.52& 45.93 / 95.31 / 24.48 & 43.22 / {\color{white}0}6.73 / 67.59  & 38.58 / 97.91 / 18.52\\
                & I-BAU \cite{zeng2022adversarial} & 42.24 / {\color{white}0}\textbf{0.05} / 70.97 & 43.27 / {\color{white}0}7.89 / 67.31  & 41.18 / 25.88 / 56.07 & 37.05 / 39.20 / 48.09 & 36.79 / {\color{white}0}5.66 / 64.91 & 36.63 / 14.74 / 59.13\\
                \midrule
                \multicolumn{4}{l}{\emph{Robust Quantization (w/ 1\% clean unlabeled data)}} \\
                & OMSE \cite{choukroun2019omse} & 56.89 / 47.07 / 54.79 & 55.72 / 22.95 / 66.01  & 54.57 / 99.27 / 26.07 & 43.96 / {\color{white}0}\textbf{0.38} / 70.96 & 43.26 / 85.11 / 28.42  & 52.13 / 91.13 / 28.69 \\
                & OCS \cite{zhao2019ocs} & 55.68 / 59.74 / 47.85 & 55.49 / 50.84 / 51.95  & 58.45 / {\color{white}0}1.01 / 77.14 & {\color{white}0}0.50 / 94.88 / {\color{white}0}1.98 & {\color{white}0}0.59 / {\color{white}0}3.44 / 47.92 & {\color{white}0}1.12 / {\color{white}0}\textbf{0.01} / 48.74 \\
                & ACIQ \cite{banner2019aciq} & 56.78 / 10.11 / 73.21 & 54.64 / 99.40 / 26.86  & 56.09 / 96.27 / 28.33 & 48.19 / 65.82 / 40.36 & 47.47 / 96.18 / 24.99 & 45.74 / 96.87 / 22.62\\
                \midrule
                & Ours  & 56.99 / {\color{white}0}0.50 / \textbf{78.12} & 55.46 / {\color{white}0}4.25 / 75.23 & 58.47 / {\color{white}0}{0.86} / \textbf{77.23} & \textbf{55.32} / {\color{white}0}2.41 / \textbf{75.63} & \textbf{54.83} / {\color{white}0}1.73 / 75.89  & \textbf{57.54} / {\color{white}0}0.62 / \textbf{76.65}\\
                \bottomrule 
            \end{tabular}
        }
    \label{tab:comparison-tiny}
\end{table*}

\vspace{0.3em}
\noindent\textbf{Backdoor Defenses and Settings.} We consider 8 possible baseline defenses, which are categorized into backdoor defenses and robust quantization. We consider 5 SOTA backdoor defenses, including FT, FP \cite{liu2018fp}, MCR \cite{zhao2020mcr}, NAD \cite{li2020nad}, and I-BAU \cite{zeng2022adversarial}. We assume all these defenses to access 5\% clean labeled data, which is their default setting. Due to the inability of quantized models to back-propagate gradients, we evaluate their effectiveness by applying them to the full-precision model and then test the model after standard quantization. All activations are also quantized to the same bandwith of weights. For robust quantization, we note that there exist many PTQ techniques but few of them have considered robustness against quantization-conditioned backdoors. Therefore, evaluations of their robustness against various conditioned backdoors are scarce and this work is to the best of our knowledge the first trial. For simplicity, we follow \citet{hong2021qu} and evaluate 3 robust quantization techniques, namely OMSE \cite{choukroun2019omse}, OCS \cite{zhao2019ocs}, and ACIQ \cite{banner2019aciq}, with 1\% clean unlabeled data provided as the calibration set. For our EFRAP, we also use 1\% clean unlabeled data, aligning with the current practice of the off-the-shelf quantization methods. We use Adam optimizer with default hyperparameters, a learning rate of 0.001, and a batch size of 32. Both $\lambda_A$ and $\lambda_P$ are set to 1. We optimize the network layer-by-layer until convergence, which takes about 7 minutes to quantize a ResNet-18 model on Tiny-ImageNet with a single NVIDIA RTX 3090 GPU. We evaluate baseline defenses on each attack setting and compare them with EFRAP. See more implementation details in \textbf{Appendix}.

\vspace{0.3em}
\noindent\textbf{Evaluation Metrics.} We involve three metrics to evaluate the performance of each baseline and our method: Attack Success Rate (\textbf{ASR}), Clean Data Accuracy (\textbf{CDA}), and Defense Trade-off Metric (\textbf{DTM}). ASR is calculated as the percentage of backdoored samples that the model incorrectly classifies into the target label. Meanwhile, CDA is computed as the proportion of correctly labeled clean samples within the test dataset. Observing that some defenses eliminate the backdoor with a notable drop in CDA, which is often unacceptable in real-world cases, DTM is first proposed in this work to measure the overall competitiveness of different backdoor defenses under the same setting. DTM considers both ASR and CDA, and it is calculated as:
\begin{equation}
\text{DTM} = (1-\alpha)\cdot\text{CDA} - \alpha\cdot\Delta\text{ASR},
\end{equation}
where $\Delta\text{ASR}$ is the difference of ASR before and after defense. Here, $\alpha$ is a weighting parameter ranging between 0 and 1. A smaller $\alpha$ value means more emphasis on CDA while a larger $\alpha$ value indicates the decrease of ASR is more critical. We select $\alpha = 0.5$ that equally weights ASR and CDA. DTM $\in [0,1]$ measures the defense's trade-off between CDA and ASR. A high DTM means the model after defense maintains a high CDA (or even increases) while eliminating backdoor effects well, while a low DTM means the defense cannot clean the backdoor well or suffers some trade-off in CDA. For example, a defense that incurs an $x\%$ decrease in ASR at the cost of an $x\%$ decrease in CDA will result in no change in the DTM. A successful defense is expected to have high CDA ($\uparrow$), low ASR ($\downarrow$), and high DTM ($\uparrow$). We repeat each experiment at least 3 times (with different random seeds) and report averaged results. In evaluating ASR, we exclude samples whose labels already belong to the target class of the attack to ensure a fair comparison.

\begin{figure*}[h]
  \begin{subfigure}[b]{0.235\textwidth}
    \includegraphics[width=\textwidth]{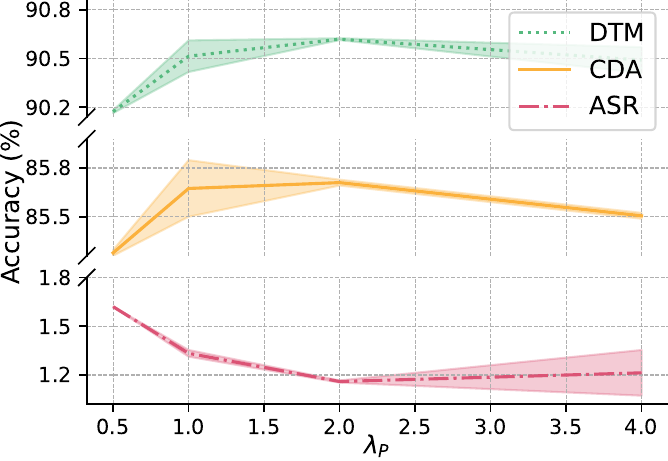}
    \caption{$\lambda_A=0.5$}
  \end{subfigure}
  \hfill
  \begin{subfigure}[b]{0.235\textwidth}
    \includegraphics[width=\textwidth]{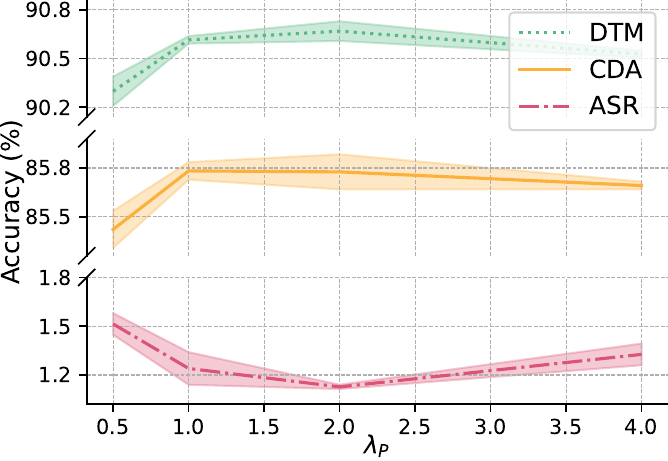}
    \caption{$\lambda_A=1$}
  \end{subfigure}
  \hfill
  \begin{subfigure}[b]{0.235\textwidth}
    \includegraphics[width=\textwidth]{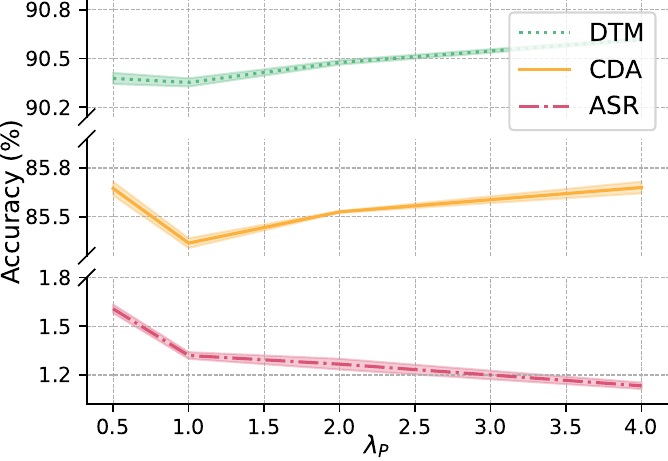}
    \caption{$\lambda_A=2$}
  \end{subfigure}
  \hfill
  \begin{subfigure}[b]{0.235\textwidth}
    \includegraphics[width=\textwidth]{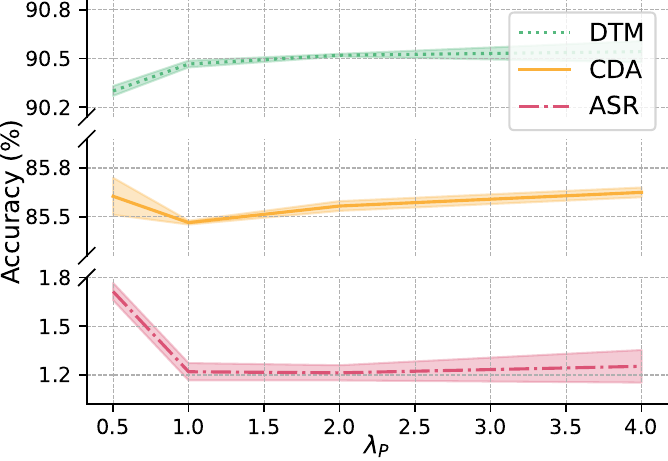}
    \caption{$\lambda_A=4$}
  \end{subfigure}
  \vspace{-5pt}
  \caption{\textbf{Ablation study on weighting parameters.} We repeat each experiment three times.}
  \label{fig:ablation}
  \vspace{-15pt}
\end{figure*}
\subsection{Experimental Results}
\label{sec:expr}
\noindent\textbf{Main Results.} The main experimental results are in Table \ref{tab:comparison-cifar10} and Table \ref{tab:comparison-tiny}. With only 1\% clean unlabeled data, EFRAP achieves the best result or nearly the best result among all baselines, across all datasets and attack settings, on all evaluation metrics. In contrast, the SOTA backdoor defenses, though provided with more data and label notations, either totally failed in handling these sneaky conditioned backdoors or performed vary from case to case. For example, on CIFAR10 dataset, FT, MCR and I-BAU achieved promising results on CompArtifact and Qu-Anti-zation, but all failed to defend against the advanced PQBackdoor; NAD can reduce the backdoor effect on PQBackdoor but severely harms clean accuracy ($\sim$ 40\% decrease on CDA), making it an infeasible defense, as indicated by a low DTM; The performance of FP is also intriguing: it often preserves CDA well, but almost failed to remove any backdoor effect on CIFAR10 dataset, which is also indicated by a consistently low DTM though it has the best CDA in some cases. Interestingly, it can mitigate the backdoor effect well for PQBackdoor on Tiny-ImageNet, at the cost of nearly 10\% CDA drop, while other defenses mostly failed. In terms of robust quantization, there also exists no encouraging defense results, with fluctuating ASR, unstable CDA, and low DTM in different settings. OCS is also observed to totally destroy the network in some cases, which is not typically the case on models without backdoors. To summarize, all existing backdoor defenses and robust quantization are inadequate in handling the intractable quantization-conditioned backdoors, while the proposed method shows robustness against all attacks across different settings, with a remarkably high CDA, DTM, and consistently low ASR.

As a final remark, an interesting observation is that certain defenses, notably MCR and our approach, can enhance CDA in ways not typically observed in conventional attacks and defenses. A possible explanation is a quantized model (especially in low bits) has only limited capacity to handle different tasks and the backdoor task occupies some of it, therefore harming CDA. When the backdoor is removed, the capacity of the quantized model can be fully utilized by the main task, resulting in notable increases in CDA. We leave a more in-depth investigation to future work.

\vspace{3pt}
\noindent\textbf{Effectiveness across Models Architectures.} We evaluate EFRAP across different model architectures, including AlexNet \cite{krizhevsky2012imagenet}, VGG-16 \cite{simonyan2014very} and MobileNet-V2 \cite{sandler2018mobilenetv2}. As shown in Table \ref{tab:results-models}, EFRAP consistently eliminates backdoor effects well while preserving high benign accuracy, demonstrating its robustness across different models.

\begin{figure}[t]
    \centering
    \includegraphics[width=\linewidth]{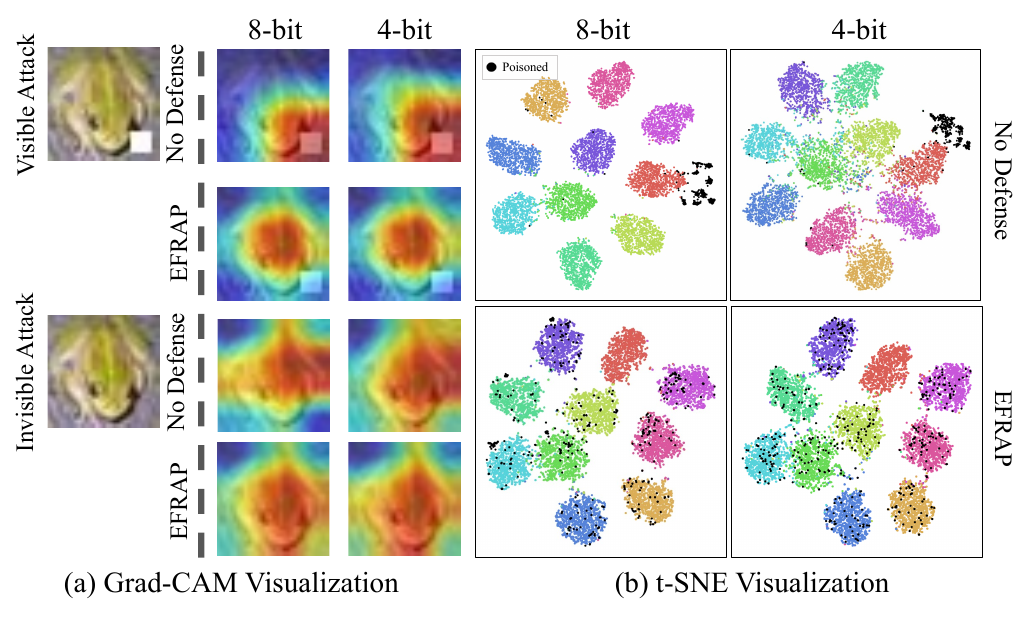}
    \vspace{-20pt}
    \caption{\textbf{Visualization results.} Grad-CAM \cite{selvaraju2017grad} highlights areas in images crucial for DNN's decisions and t-SNE \cite{van2008visualizing} visualizes data in a DNN's low-dimensional feature space. The model is ResNet-18 and the dataset is CIFAR-10.}
    \label{fig:gradcam}
    \vspace{-15pt}
\end{figure}

\vspace{3pt}
\noindent\textbf{Grad-CAM \cite{selvaraju2017grad} and t-SNE \cite{van2008visualizing} Visualizations.} These methods are widely used to interpret model predictions. We train models attacked by \cite{hong2021qu} and \cite{ma2023commercial} with visible patch-based triggers \cite{gu2017badnets} and invisible triggers \cite{nguyen2020wanet}. We visualize the Grad-CAM results on images before and after defense and visualize the attacked model of \cite{ma2023commercial} using t-SNE. As shown in Figure \ref{fig:gradcam}, Grad-CAM results of defended models focus on the image's subject rather than trigger regions as in backdoored ones, and t-SNE shows post-defense dispersion of poisoned samples, rather than clustering. These results indicate that backdoors are indeed successfully removed.

\begin{table}[h]
    \centering
    \footnotesize
    \setlength{\tabcolsep}{4pt}
    \caption{\textbf{Defense results across different models.} We evaluate EFRAP against 4-bit attack \cite{hong2021qu} on CIFAR-10. }
    \vspace{-0.8em}
         \scalebox{0.9}{
         \begin{tabular}{c c c}
         \toprule
              Models & Defense & CDA $\uparrow$ / ASR $\downarrow$ / DTM $\uparrow$ \\
              \midrule
              \multirow{2}{*}{AlexNet \cite{krizhevsky2012imagenet}} & No Defense & 76.47 / 88.71 / 38.24\\
              & EFRAP & 80.58 / {\color{white}0}1.30 / 84.00 \\
              \midrule
              \multirow{2}{*}{VGG-16 \cite{simonyan2014very}} & No Defense & 82.78 / 98.57 / 41.39\\
              & EFRAP & 86.17 / {\color{white}0}1.26 / 90.05\\
              \midrule
              \multirow{2}{*}{MobileNet-V2 \cite{sandler2018mobilenetv2}} & No Defense & 79.80 / 99.90 / 39.90\\
              & EFRAP & 87.58 / {\color{white}0}1.46 / 93.01 \\
        \bottomrule
        \end{tabular}
        }
    \label{tab:results-models}
\end{table}

\subsection{Ablation Studies}
\label{sec:ablation}
All ablation studies are conducted on \cite{ma2023commercial} for both 8-bit and 4-bit settings on ResNet-18. The dataset is CIFAR10. Due to space limit, 8-bit results are placed in the \textbf{Appendix}.

\begin{table}[!t]
\vspace{-5pt}
    \centering
    \footnotesize
    \setlength{\tabcolsep}{4pt}
    \caption{\textbf{Ablation study on each component.}}
    \vspace{-0.8em}
        \centering
         \scalebox{0.9}{
         \begin{tabular}{c c c |c}
         \toprule
               \multicolumn{3}{c|}{Component} & {4-bit Attack}\\
               $\mathcal{L}_{F}$  & $\mathcal{L}_{A}$ & $\mathcal{L}_{P}$ & CDA $\uparrow$ / ASR $\downarrow$ / DTM $\uparrow$\\ 
        \hline
                 $-$ & $-$ & $-$ & 81.31 / 96.74 / 40.66\\
        \hline
                $\checkmark$ & $-$ & $-$ & 51.47 /{\color{white}0}\ 5.73 / 71.24\\
                $\checkmark$ & $\checkmark$ & $-$ & 84.36 / {\color{white}0}{1.68} / 89.71 \\
        \hline
               $\checkmark$ & $\checkmark$ & $\checkmark$   & \textbf{85.16} / {\color{white}0}2.33 / \textbf{89.79}\\
        \bottomrule
        \end{tabular}
        }
    \label{tab:ablation-component}
\end{table}

\vspace{5pt}
\noindent\textbf{Effectiveness of Each Component.} EFRAP consists of error-guided flipped rounding and activation preservation, represented by $\mathcal{L}_F$ and $\mathcal{L}_A$, respectively. We study the effectiveness of each component and the results are in Table.~\ref{tab:ablation-component}. To conclude, every component of EFRAP is indispensable, where $\mathcal{L}_F$ destroys essential backdoor connections and $\mathcal{L}_A$ compensates for CDA. Though $\mathcal{L}_P$ does not greatly influence the result, it makes training more stable. 

\vspace{5pt}
\noindent\textbf{Effect of Weighting Parameters $\lambda_A$ and $\lambda_P$.} The relative strength of $\mathcal{L}_A$ and $\mathcal{L}_P$ is controlled by the weighting parameter $\lambda_A$ and $\lambda_P$. As illustrated in Figure \ref{fig:ablation}, EFRAP is not sensitive to the choice of weighting parameters. Thus, we empirically set both of them to $1$ in our experiments.

\subsection{Resistance to Potential Adaptive Attacks}
\vspace{-5pt}
To evaluate the robustness of our EFRAP, we test its resistance against adaptive attacks. Specifically, we attack EFRAP by enforcing the dormant backdoor to be activated even if the weights are flipped rounded. Experimental results show that this attack indeed work well when all neurons are flipped (CDA=92.12\%, ASR=98.57\%). However, it failed to attack EFRAP (CDA=92.16\%,  ASR=1.74\%). The most probable reason is EFRAP flips neurons selectively based on the overall objective, rather than all. The detailed discussions are in the \textbf{Appendix}.

\vspace{-5pt}
\section{Conclusion}
\vspace{-5pt}
In this paper, for the first time, we introduce a defense against quantization-conditioned backdoor attacks that maliciously exploit standard model quantization. Through analyses of truncation errors in neuron weights, we revealed how quantization triggers dormant backdoors. Build upon this, we propose EFRAP, a method learning a non-nearest quantization rounding strategy, to counteract backdoor effects while preserving clean accuracy. Extensive evaluations and comparisons confirm the effectiveness and robustness of EFRAP. We call for more attention on DNN lifecycle security and expect future research on building effective detections and defenses for conditioned backdoor attacks.

\section*{Acknowledgements}
\vspace{-5pt}
This work was supported by the National Key Research and Development Program of China under Grant 2021YFB3100300, the National Natural Science Foundation of China under Grants U20A20178, 62072395, 62206207, 62202340, and 62372334, and the CCF-NSFOCUS `Kunpeng' Research Fund (CCF-NSFOCUS 2023005). This work was partly done when Boheng Li was a (remote) Research Intern at The State Key Laboratory of Blockchain and Data Security, Zhejiang University.

{
    \small
    \bibliographystyle{ieeenat_fullname}
    \bibliography{main}
}

\input{supp}

\end{document}

%% file: preamble.tex
\usepackage[dvipsnames]{xcolor}


%% file: supp.tex
\appendix
\twocolumn[{%
 \noindent
 \LARGE Supplementary Material\\[0.5em]
 \large Nearest Is Not Dearest: Towards Practical Defense against Quantization-conditioned Backdoor Attacks\\[1em]
}]

\section{More Implementation Details}
\label{sec:implement}

\vspace{5pt}
\noindent\textbf{More Details on Datasets.} In the main paper, we evaluate EFRAP and compare it with baseline defenses on 2 benchmarking datasets: CIFAR10 \cite{krizhevsky2009learning} and Tiny-ImageNet \cite{le2015tiny}. In this supplementary material, we also evaluate EFRAP on a high-resolution dataset, \ie, ImageNette \cite{imagenette_dataset}. Here is a brief introduction for each of them:

\begin{itemize}
    \item \textbf{CIFAR-10 \cite{krizhevsky2009learning}}: This dataset, originating from the Canadian Institute For Advanced Research, comprises 60,000 color images of 32$\times$32 pixels, spread across 10 different classes, with 6,000 images per class. It includes a split of 50,000 training and 10,000 test images, making it a staple in research for image classification tasks.
    \item \textbf{Tiny-ImageNet \cite{le2015tiny}}: Comprising 100,000 images downsized to 64$\times$64 pixels, Tiny-ImageNet is structured into 200 classes, each with 500 training, 50 validation, and 50 test images. This dataset serves as a compact version of ImageNet, catering to visual recognition challenges.
    \item \textbf{ImageNette \cite{imagenette_dataset}}: This dataset is a subset of ImageNet, widely used in the research community \cite{xiang2021patchguard,zhang2024why}. It consists of 9,469 training images and 3,925 test images. Each image is with a high resolution of 224$\times$224.
\end{itemize}

These datasets are widely used to evaluate backdoor attacks' performances on DNNs for computer vision \cite{dong2023mind,NEURIPS2022_db1d5c63,nguyen2023iba} and are also the benchmarking datasets for SOTA backdoor benchmarks and toolboxes \cite{li2023backdoorbox,wu2022backdoorbench}.

\vspace{5pt}
\noindent\textbf{More Details on Backdoor Attacks.} 
In this work, we evaluate 3 existing quantization-conditioned backdoor attacks, \ie, CompArtifact \cite{tian2022stealthy}, Qu-ANTI-zation \cite{hong2021qu}, and PQBackdoor \cite{ma2021quantization,ma2023commercial}. Below is the introduction of each attack and their implementation details:
\begin{itemize}
    \item \textbf{CompArtifact \cite{tian2022stealthy}}: CompArtifact uses the trigger pattern from BadNets \cite{gu2017badnets}, \ie, a 3$\times$3 small white patch on the right lower corner of the image. It is robust to calibration set changes but has low transferability across different bandwidths. Therefore, in our work, we respectively train compromised models for each bandwidth for fair comparison. We use their released official code\footnote{https://github.com/yulongt23/Stealthy-Backdoors-as-Compression-Artifacts}. Following their original design, we first train a clean model for 400 epochs using standard cross-entropy loss, and then re-train each model (respectively for 8-bit and 4-bit) with the modified objective for 50 epochs, where the poison rate is set to $50\%$ during re-training.
    \item \textbf{Qu-ANTI-zation \cite{hong2021qu}}: To help the attack transfer, Qu-ANTI-zation considers multiple bit bandwidths in the re-training stage. It showed robustness against several quantization bandwidths as well as robust quantization techniques. It also uses the patch-based trigger, whereas the size is set to 4$\times$4 on CIFAR10 and 8$\times$8 on Tiny-ImageNet. In our evaluation, we use their released official code\footnote{https://github.com/Secure-AI-Systems-Group/Qu-ANTI-zation}. Following their original design, we first train a clean model for 200 epochs. Then we re-train the model with the modified objective for 50 epochs, where the poison rate is also set to $50\%$ during re-training. 
    \item \textbf{PQBackdoor \cite{ma2021quantization,ma2023commercial}}: PQBackdoor is the most recent and the SOTA quantization-conditioned backdoor attack. It improves the training pipeline via introducing a two-stage attack strategy: firstly, train a backdoored full-precision model, and secondly, make the backdoor dormant by re-training using the projected gradient descent \cite{madry2018towards}. This stabilizes the training of the quantization-conditioned backdoor and further improves its robustness. It also uses the patch-based trigger and the size is set to 6$\times$6. PQbackdoor also demonstrated its robustness against blind backdoor defenses such as fine-tuning, and its transferability to commercial quantization frameworks like PyTorch Mobile \cite{paszke2019pytorch} and TensorFlow Lite \cite{abadi2016tensorflow}. We use the official PyTorch source code from the authors\footnote{https://github.com/quantization-backdoor} and follow their settings in the paper. For the first stage, the poisoning rate is set to $1\%$, with the standard training pipeline on poisoning-based backdoor attacks for 100 epochs. After the first stage, the poisoning rate is then set to $50\%$ in the second stage, which takes another 50 epochs. Unfortunately, even if we tried several times (>5), we failed to obtain a full-precision model with CDA reported in their paper. On  CIFAR10, we can only have 86.43\% with ResNet-18 during our reproduction, much lower than 93.44\% reported in their original paper. On Tiny-ImageNet, the CDA is even worse (35.5\%), which is much lower than the clean models (usually around 58\%). This makes the attacked model less likely to be used by the victim. A possible reason is the network does not fully converge during the first stage (only 100 epochs). To verify this, we train another model for 400 epochs during the first stage and find we can indeed obtain a model with higher precision (93.03\% on CIFAR10 and 58.5\% on Tiny-ImageNet). To best align with the paper setting and consider the real-world scenarios, in our main paper, we report the results of PQBackdoor with these lower CDA models on CIFAR10 but higher CDA models on Tiny-ImageNet. We place the defense results for the higher CDA model in Table \ref{tab:higher-CDA}.
\end{itemize}

\begin{table}[t]
    \centering
    \footnotesize
    \setlength{\tabcolsep}{4pt}
    \caption{\textbf{Defense Results on PQBackdoor (Higher CDA Model) on CIFAR10 (\%).} Standard means standard quantization.}
         \scalebox{0.9}{
         \begin{tabular}{c c c}
         \toprule
              Bandwidth & Setting &  CDA / ASR \\
              \midrule
              32-bit & Full-precision & 93.02 / {\color{white}0}9.20 \\
              \midrule
              \multirow{2}{*}{8-bit} & Standard & 92.49 / 97.43 \\
              & EFRAP & 90.49 / {\color{white}0}7.28 \\
              \midrule
              \multirow{2}{*}{4-bit} & Standard & 89.79 / 96.49 \\
              & EFRAP & 89.80 / {\color{white}0}0.64 \\
        \bottomrule
        \end{tabular}
        }
    \label{tab:higher-CDA}
    \vspace{-10pt}
\end{table}

\vspace{5pt}
\noindent\textbf{More Details on Backdoor Defenses.} In this paper, we considered 8 possible defenses against quantization-conditioned backdoor attacks, which are broadly classified into backdoor defense and robust quantization. For backdoor defenses, we consider 5 SOTA backdoor defenses, including FT \cite{sha2022fine}, FP \cite{liu2018fp}, MCR \cite{zhao2020mcr}, NAD \cite{li2020nad}, and I-BAU \cite{zeng2022adversarial}. For all defenses, we use the open-source code from BackdoorBox\footnote{https://github.com/THUYimingLi/BackdoorBox} \cite{li2023backdoorbox}, except for I-BAU, which we use their official implementation\footnote{https://github.com/YiZeng623/I-BAU}. Here are their brief introduction and implementation details:

\begin{table}[t]
    \centering
    \footnotesize
    \setlength{\tabcolsep}{4pt}
    \caption{\textbf{Results on Full-precision Models (\%).} For CompArtifact, we train 2 models, respectively for 8-bit attack (8-bit) and 4-bit attack (4-bit). For PQBackdoor (higher CDA), the backdoor model is trained with 400 epochs during the first stage.}
         \scalebox{0.9}{
         \begin{tabular}{c c c}
         \toprule
              Dataset & Attack & CDA / ASR \\
              \midrule
              \multirow{6}{*}{CIFAR10} & Clean Model & 93.44\% / 0.44\%\\
              & CompArtifact (8-bit) & 91.46\% / 1.26\%\\
              & CompArtifact (4-bit) & 93.68\% / 1.33\%\\
              & Qu-ANTI-zation & 93.17\% / 2.18\%\\
              & PQBackdoor & 86.43\% / 2.67\% \\
              & PQBackdoor (higher CDA) & 93.02\% / 3.20\% \\
              \midrule
              \multirow{5}{*}{Tiny-ImageNet} & Clean Model & 57.77\% / 0.21\%\\
              & CompArtifact (8-bit) & 57.09\% / 0.78\%\\
              & CompArtifact (4-bit) & 56.89\% / 1.43\%\\
              & Qu-ANTI-zation & 55.82\% / 2.16\%\\
              & PQBackdoor (higher CDA) & 58.50\% / 0.88\% \\
        \bottomrule
        \end{tabular}
        }
    \label{tab:full-precision}
\end{table}

\begin{itemize}
    \item \textbf{FT \cite{sha2022fine}:} Fine-tuning (FT) is the most frequently considered baseline for backdoor defenses. It directly fine-tunes the model using a small set of clean data. Though sounds simple, it can effectively remove backdoor effects for many SOTA backdoor attacks \cite{wu2022backdoorbench}. In our work, we fine-tune all layers of the compromised full-precision model using 5\% clean data for 50 epochs.
    \item  \textbf{FP \cite{liu2018fp}:} Fine-pruning (FP) is a defense combining fine-tuning and pruning. It first feeds a small set of clean data to the network and measures the activation, then prunes the neurons less frequently activated (which are considered backdoor neurons). To maintain clean accuracy, fine-tuning is involved after pruning.  In our work, we measure the activation of the last residual block and the pruning rate is set to 0.4. We then fine-tune the model with 5\% clean data for 50 epochs.
    \item \textbf{MCR \cite{zhao2020mcr}:} Mode connectivity repair (MCR) is a defense that visits DNN life-cycle security from the loss landscape's perspective. It first fine-tunes a backdoored model, then employs mode connectivity in loss landscapes between the original backdoored model and the fine-tuned model, and finally measures and removes backdoor functions through mode connectivity repair. In our work, we first fine-tune the backdoored model for 50 epochs, then run 100 epochs of curvenet training, and finally 100 epochs of model updating. The hyperparameter $t$ is respectively set to 0.1 and 0.9 and we report the results with higher DTM.
    \item \textbf{NAD \cite{li2020nad}:} Neural attention distillation (NAD) is a defense using knowledge distillation with attention guidance. It observes that the attention of backdoored and clean models are different, so it first fine-tunes a backdoored model, which is referred to as a less poisonous model, and then uses this less poisonous model as the teacher model, the original backdoored model as the student model, and conduct knowledge distillation with attention alignment guidance. We run 50 epochs of fine-tuning to obtain the teacher model and 50 epochs to purify the student model.
    \item \textbf{I-BAU \cite{zeng2022adversarial}:} Implicit backdoor adversarial unlearning (I-BAU) views the task of backdoor removal as a minimax formulation. It then utilizes the implicit hypergradient to account for the interdependence between inner and outer optimization. It is shown faster, more computationally efficient, and more effective than previous backdoor defenses, achieving SOTA defense results on many benchmarks \cite{wu2022backdoorbench}. We run 3 rounds of I-BAU for each attack.
\end{itemize}

All the aforementioned backdoor defenses have shown effectiveness against SOTA backdoor attacks \cite{wu2022backdoorbench,li2023backdoorbox}, not to mention the rudimentary backdoor of BadNets \cite{gu2017badnets} used by many quantization-conditioned backdoors. As reported in \cite{wu2022backdoorbench}, many evaluated defenses in this paper can reduce the ASR of BadNets to nearly 0\% while maintaining high clean accuracy. However, as we show in the main paper, their performances are largely weakened or even ineffective. The main possible reason is that these conditioned backdoors stay dormant on the full-precision models, making the assumption of many backdoor defenses (assuming the existence of explicit backdoors) invalid.

For robust quantization, following \citet{hong2021qu}, OMSE \cite{choukroun2019omse}, OCS \cite{zhao2019ocs}, and ACIQ \cite{banner2019aciq} are considered. Here are their brief introduction and implementation details:

\begin{itemize}
    \item \textbf{OMSE \cite{choukroun2019omse}:} Optimal MSE (OMSE) is a widely used technique for robust post-training quantization. It formalizes the linear quantization task as a Minimum Mean Squared problem for both weights and activations and solves it via layer-wise optimization. It can largely avoid the severe expected behavioral change of vanilla quantization.  
    \item \textbf{OCS \cite{zhao2019ocs}:} Outlier channel splitting (OCS) improves quantization via duplicating channels containing outliers and halving the channel values, thus largely avoiding outliers in the distribution. It is shown superior than SOTA clipping techniques with only minor overhead.
    \item \textbf{ACIQ \cite{banner2019aciq}:} Analytical clipping for integer quantization (ACIQ) analytically computes the clipping range as well as the per-channel bit allocation for DNNs, thus enhancing the robustness of model quantization.
\end{itemize}

\vspace{5pt}
\noindent\textbf{Implementation Details.} For all experiments, we use Python 3.8.18 and PyTorch 1.10.0+cu113 framework, with torchvision 0.11.1. All experiments are implemented in Python and run on a 14-core Intel(R) Xeon(R) Gold 5117 CPU @2.00GHz with a single NVIDIA GeForce RTX 3090 GPU machine running Linux version 5.4.0-144-generic (buildd@lcy02-amd64-089) (Ubuntu 9.4.0-1 ubuntu1\~20.04.1). Unless otherwise stated, we use Adam optimizer \cite{kingma2014adam} with default parameters. All other hyperparameters follow the original setting described in the paper. During clean model training and backdoor model training (first stage for PQBackdoor), the learning rate is set to 1e-3, whereas it is set to 1e-4 for all backdoor defenses and the second stage of PQBackdoor. The batch size is set to 64 for CIFAR10 and Tiny-ImageNet, and 16 for ImageNette. Each attack finally results in a full-precision model with a dormant backdoor inserted on each dataset and model architecture. For all experiments, we repeat the experiment at least three times and report the average results in the paper. The standard deviation are small (usually less than $2\%$ for both ASR and CDA). Unless otherwise stated, all activations are also quantized with the same bandwidth of weights. As shown in Table \ref{tab:full-precision}, quantization-conditioned backdoors hide well on full-precision models, with a CDA similar to that of a clean model, and an ASR of nearly $0\%$.

\begin{figure*}[ht]
  \centering
  \begin{subfigure}[b]{0.24\textwidth}
    \includegraphics[width=\textwidth]{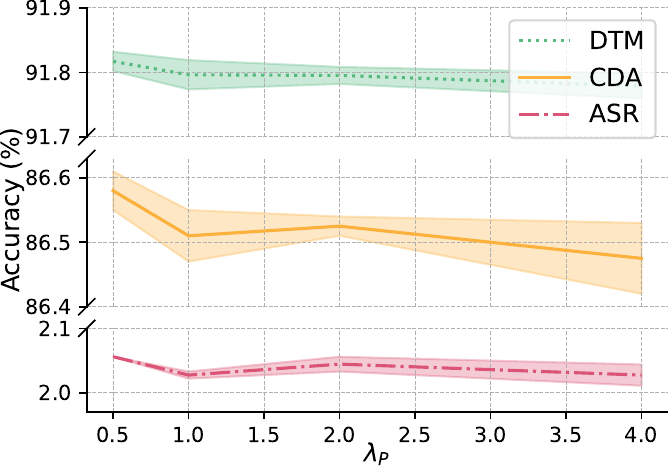}
    \caption{8-bit, $\lambda_A=0.5$}
  \end{subfigure}
  \hfill
  \begin{subfigure}[b]{0.24\textwidth}
    \includegraphics[width=\textwidth]{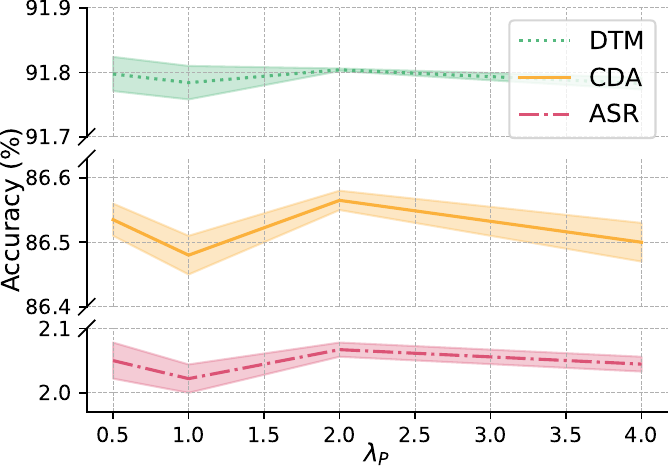}
    \caption{8-bit, $\lambda_A=1$}
  \end{subfigure}
  \hfill
  \begin{subfigure}[b]{0.24\textwidth}
    \includegraphics[width=\textwidth]{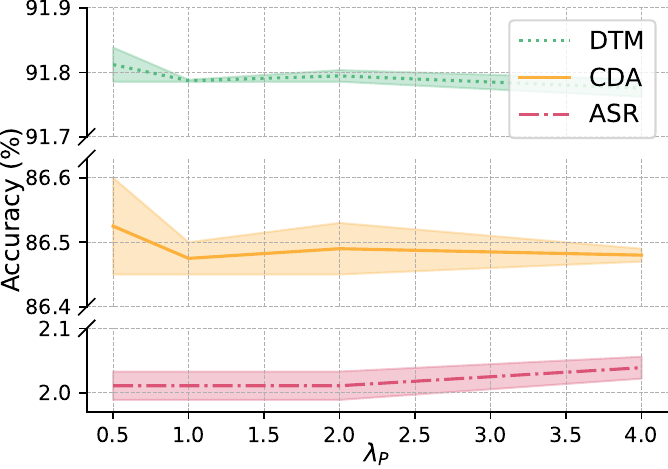}
    \caption{8-bit, $\lambda_A=2$}
  \end{subfigure}
  \hfill
  \begin{subfigure}[b]{0.24\textwidth}
    \includegraphics[width=\textwidth]{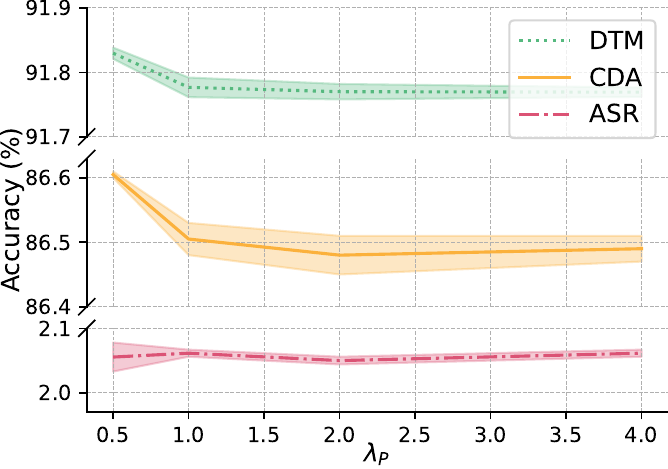}
    \caption{8-bit, $\lambda_A=4$}
  \end{subfigure}
  \caption{\textbf{Ablation Study on Weighting Parameters.} We repeat each experiment three times.}
  \label{fig:ablation}
\end{figure*}

\begin{figure*}[ht]
  \centering
  \begin{subfigure}[b]{0.49\textwidth}
    \includegraphics[width=0.49\textwidth]{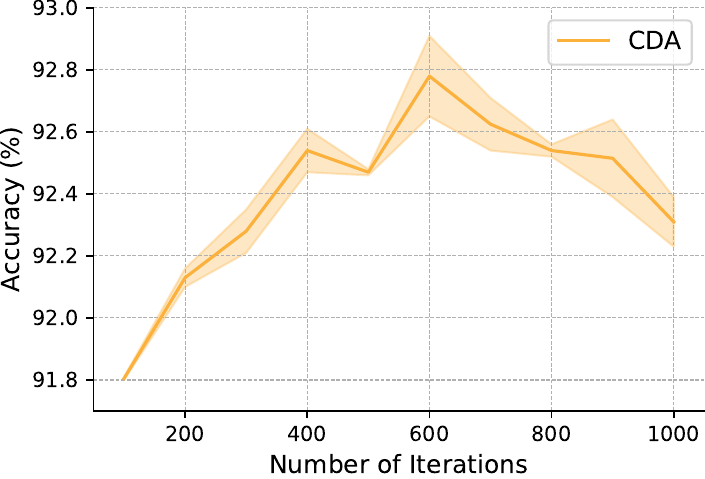}
    \hfill
    \includegraphics[width=0.49\textwidth]{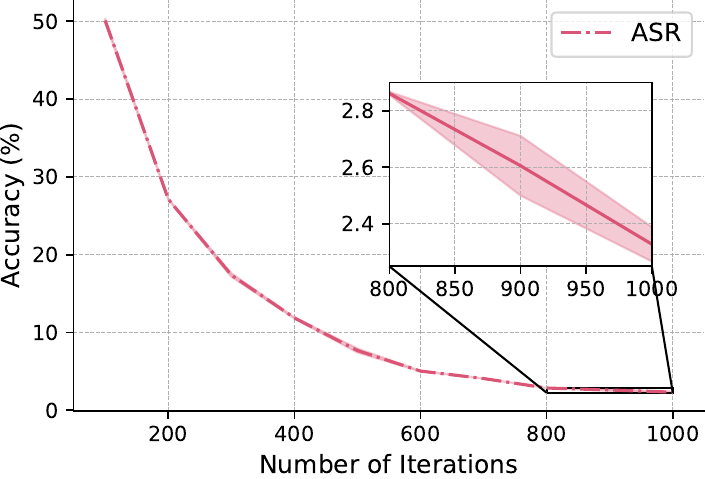}
    \caption{8-bit Attack Results}
  \end{subfigure}
  \hfill
  \begin{subfigure}[b]{0.49\textwidth}
    \includegraphics[width=0.49\textwidth]{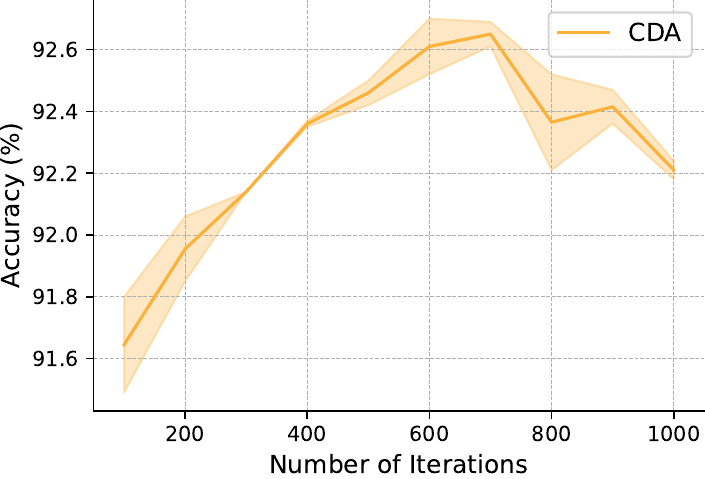}
    \hfill
    \includegraphics[width=0.49\textwidth]{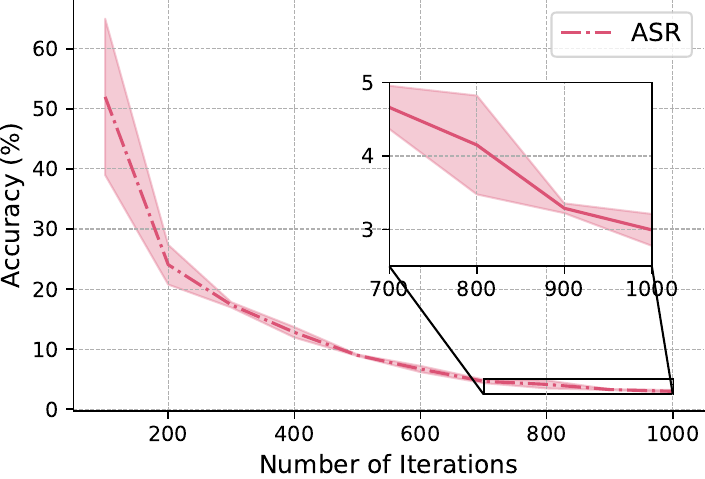}
    \caption{4-bit Attack Results}
  \end{subfigure}
  \caption{\textbf{Ablation Study on Number of Iterations.} We repeat each experiment three times.}
  \label{fig:num-iter}
\end{figure*}

\section{More Ablation Studies}
\subsection{Ablation Study on 8-bit Attacks}

Due to the page limit, the ablation study on 8-bit attacks is placed in the Appendix. Except for the evaluated bandwidth, other settings are the same as in the main paper. Here are the ablation results:

\vspace{5pt}
\noindent\textbf{Effectiveness of Each Component.} As shown in Table \ref{tab:ablation-component}, on 8-bit attacks, the results keep a similar trend as in 4-bit attacks. Different from 4-bit attacks, the $\mathcal{L}_F$ term alone does not cause severe harm to CDA. This is probably because the 8-bit quantization errors are small and the model learns to be robust to such small flipped rounding errors. However, we can still see that $\mathcal{L}_{A}$ restores some of the neurons critical for CDA. This further validates the effectiveness of each component proposed in EFRAP.

\vspace{5pt}
\noindent\textbf{Effect of Weighting Parameters $\lambda_A$ and $\lambda_P$.} As illustrated in Figure \ref{fig:ablation}, on 8-bit attacks, EFRAP is still not sensitive to the choice of weighting parameters on 8-bit settings. This aligns with our conclusion in the main paper.

\begin{table}[h]
    \centering
    \footnotesize
    \setlength{\tabcolsep}{4pt}
    \caption{\textbf{Ablation Study on Each Component.} $\mathcal{L}_F^\prime$ means $\mathcal{L}_F$ w/o error guidance, \ie do not multiply $E$ when calculating $\mathcal{L}_F$.}
         \scalebox{1}{
         \begin{tabular}{c c c c |c}
         \toprule
               \multicolumn{4}{c|}{Component} & {8bit Attack}\\
                $\mathcal{L}_F^\prime$ & $\mathcal{L}_{F}$  & $\mathcal{L}_{A}$ & $\mathcal{L}_{P}$ & CDA $\uparrow$ / ASR $\downarrow$ / DTM $\uparrow$\\ 
        \hline
                $-$ & $-$ & $-$ & $-$ & 85.16 / 99.11 / 42.58\\
        \hline
                $-$ & $\checkmark$ & $-$ & $-$ & 83.17 / {\color{white}0}\textbf{1.41} / 90.44\\
                $-$ & $\checkmark$ & $\checkmark$ & $-$ & 86.15 / {\color{white}0}2.03 / 91.62 \\
               $\checkmark$ & $-$ & $\checkmark$ & $\checkmark$  &  86.06 / {\color{white}0}2.99 / 91.09\\
        \hline
               $-$ & $\checkmark$ & $\checkmark$ & $\checkmark$  &  \textbf{86.52} / {\color{white}0}2.38 / \textbf{91.63}\\
        \bottomrule
        \end{tabular}
        }
    \label{tab:ablation-component}
    \vspace{-13pt}
\end{table}

\subsection{Ablation Study on Numbers of Iterations}
EFRAP optimizes the network layer-by-layer. To better understand the convergence of EFRAP, we examine the influence of the number of iterations by changing the optimization iteration of EFRAP in the layer-wise optimization. The results are shown in Figure \ref{fig:num-iter}. We can see that the attack takes effect (ASR<20\%) when iteration is above 200, and EFRAP is about to converge with 1000 iterations. To ensure convergence we uniformly take 10000 iterations for our evaluations. This takes about 7 minutes to quantize a ResNet-18 model on Tiny-ImageNet.

\section{Resistance to Potential Adaptive Attacks}
To comprehensively evaluate the robustness of EFRAP, we consider a very smart attacker who is informed of the design of EFRAP and tries to bypass it. According to our threat model, the attacker controls the total training procedure. Thus, he/she can modify the training objective, in order to bypass EFRAP. Specifically, we consider bypassing EFRAP via enforcing the conditioned backdoor to still be activated even if all neurons are flipped rounded. To facilitate a better understanding, in this section, we first present the threat model and the adaptive attack strategy. Then we analyse the effectiveness of the proposed adaptive attacks and give further discussions.

\subsection{Threat Model}
The threat model for the defender is the same as that of the main paper. As for the attacker, we assume he/she can not only control the whole training dataset as well as the training procedure but is also informed of the design of EFRAP. The attacker aims to bypass the defense via adaptive strategies.

\begin{algorithm}[t]
\caption{Adaptive Attacks (Re-training Stage)}
\small
\begin{algorithmic}[1]
\Statex \textbf{Input:} A pre-trained clean model $f$ with weights $\bm{W}$, training set $\mathcal{D}$, quantization scale $s$, learning rate $\tau$.
\Statex \textbf{Output:} Backdoored model weights $\bm{W}$.
\While{not converged}
    \Statex ~~~~~~\textcolor{gray}{$\triangleright$ \textit{Record rounding strategy of nearest rounding}}
    \State $R(\bm{W}) \leftarrow  \mathbbm{1}\{{s} \cdot \lfloor{\frac{\bm{W}}{{s}}}\rceil - \bm{W}\succ0$\} \Comment{Rounding strategy}
    \State $\overline{R}(\bm{W}) \leftarrow 1-R(\bm{W})$ \Comment{Flipped rounding strategy}
    \Statex \textcolor{gray}{$\triangleright$ \textit{Quantization with original and flipped strategy.}}
    \State $Q({\bm{W}}) \leftarrow  {s} \cdot clip \left(\bigg\lfloor{\frac{\bm{W}}{{s}}}\bigg\rfloor + {R}(\bm{W}), {n}, {p}\right)$
    \State $\overline{Q}({\bm{W}}) \leftarrow  {s} \cdot clip \left(\bigg\lfloor{\frac{\bm{W}}{{s}}}\bigg\rfloor + \overline{R}(\bm{W}), {n}, {p}\right)$
    \Statex \textcolor{gray}{$\triangleright$ \textit{Behave normally on full-precision model.}}
    \State $\mathcal{L}_{C} \leftarrow \mathcal{L}_{ce}(f(\bm{x}),y)+\alpha\cdot\mathcal{L}_{ce}(f(\bm{x_t}),y)$
    \Statex \textcolor{gray}{$\triangleright$ \textit{Backdoor objectives on nearest rounding.}}
    \State $\mathcal{L}_{Q} \leftarrow \beta\cdot\mathcal{L}_{ce}(f_Q(\bm{x}),{y})+\gamma\cdot\mathcal{L}_{ce}(f_Q(\bm{x_t}),{y_t})$
    \Statex \textcolor{gray}{$\triangleright$ \textit{Backdoor objectives on flipped rounding.}}
    \State $\mathcal{L}_{F} \leftarrow \zeta\cdot\mathcal{L}_{ce}(f_{\overline{Q}}(\bm{x}),{y})+\eta\cdot\mathcal{L}_{ce}(f_{\overline{Q}}(\bm{x_t}),{y_t})$
    \State Get a batch of $\bm{x}$ from $\mathcal{D}$
    \State $\mathcal{L} \leftarrow \mathcal{L}_C + \lambda_Q\mathcal{L}_Q + \lambda_F\mathcal{L}_F$
    \State Update $\bm{W} \leftarrow \bm{W} - \tau\cdot\nabla_{f} \mathcal{L}$
\EndWhile

\State \Return{$\bm{W}$}
\end{algorithmic}
\label{alg:adaptive}
\end{algorithm}

\subsection{Attack Methods and Results}

\noindent\textbf{Attack Method.} EFRAP's effectiveness against backdoors largely relies on the flipped rounding objective, which breaks the connection between rounding errors and backdoor activation. Therefore, an adaptive strategy is to maintain such connection when neurons are flipped, \ie, ensure the backdoor will still be activated even if all neurons are flipped. Specifically, the attacker may implement the adaptive attack by involving a new objective, \ie, maintain backdoor effects on the flipped rounded quantized model.

\noindent\textbf{Experimental Settings.} We adopt the training procedure in \cite{hong2021qu} to implement the adaptive attacks. We first train a clean full-precision model for 400 epochs, then we use the modified training pipeline in Algorithm \ref{alg:adaptive} to re-train the clean model for 50 epochs to insert the quantization-conditioned backdoor. We also tried the training procedure of PQBackdoor \cite{ma2021quantization,ma2023commercial} (first train a backdoored full-precision model then hide the backdoor using the modified objective with PGD) but the results are similar. All experiments are conducted on CIFAR10 and ResNet-18. The other hyper-parameters and implementation details follow the settings described in Section \ref{sec:implement}.

\noindent\textbf{Results \& Analysis.} The results of adaptive attacks are in Table \ref{tab:universal-attack}. On 8-bit attacks, we can see that the attack indeed works well on full-precision models and quantized models. The backdoor hides well in full-precision mode, with a high CDA and low ASR. The adaptive strategy also works well on both standard and flipped rounding strategies, with both high CDA and ASR. However, it fails to defeat EFRAP, where the defended model expresses high CDA and very low ASR. The reason is that EFRAP selectively flips the neurons based on the two objectives, rather than flipping all the neurons. We calculated the ratio of neurons flipped by EFRAP in each layer of a given model and found that the flip rates are varying from model to model and layer to layer, usually between $1\%\sim40\%$. Therefore, the final rounding strategy of EFRAP is neither nearest rounding nor flipped rounding, making it still effective in breaking the connections between rounding errors and backdoor activations. We also observe the attack results are less satisfactory (CDA=$41.36\%$ on Flipped) on the 4-bit setting. This is because the flipped rounding causes larger errors than nearest rounding, especially in the 4-bit setting, which makes it harder to maintain a high CDA.

\begin{table}[t]
    \centering
    \footnotesize
    \setlength{\tabcolsep}{4pt}
    \caption{\textbf{Results on Adaptive Attacks (\%).} Standard means standard quantization and Flipped means quantization with flipped rounding strategy.}
         \scalebox{1}{
         \begin{tabular}{c c c}
         \toprule
              Bandwidth & Setting &  CDA ~/~ ASR \\
              \midrule
              \multirow{4}{*}{8-bit} & Full-precision & 93.29 / {\color{white}0}0.84 \\
               & Standard & 93.36 / 100.0\\
              & Flipped & 92.12 / 98.57  \\
              & EFRAP & 92.16 / {\color{white}0}1.74 \\
              \midrule
              \multirow{4}{*}{4-bit} & Full-precision & 93.29 / {\color{white}0}0.84  \\
               & Standard  &   88.42 / 100.0\\ 
              & Flipped & 41.36 / 99.88 \\
              & EFRAP & 92.35 / {\color{white}0}1.12 \\
        \bottomrule
        \end{tabular}
        }
    \label{tab:universal-attack}
    \vspace{-20pt}
\end{table}

\noindent\textbf{More Advanced Attacks.} Considering the failure of directly implanting backdoors into the flipped rounding strategy, we consider two more advanced adaptive attacks: random flipping and adversarial training with EFRAP. Random flipping refers to randomly flipping some neurons' rounding strategy at each iteration, while the adversarial training with EFRAP refers to conducting EFRAP every single iteration and implanting backdoors into the rounding strategy of EFRAP. These two strategies simulate the possible effect of EFRAP and expect to learn a robust backdoor against it. Note that adversarial training with EFRAP is very time-consuming as conducting EFRAP each time requires about 7 minutes. Therefore, it takes about $50\times781\times7/60=4555$ GPU hours or nearly 190 GPU days to re-train a single ResNet-18 model on CIFAR10 for a re-training stage of 50 epochs, in stark contrast to the original re-training, which takes only 1.5 hours. Therefore, we conduct EFRAP every 50 steps, and the iteration of EFRAP is set to 1000 as a computationally feasible proxy. However, both these strategies failed to bypass EFRAP, even though we tried different flip rates (from $1\%$ to $40\%$), learning rates, batch size, \etc, for several times. These attacks all either fail to defeat EFRAP (with a high CDA and very low ASR), or the network can only get bad performances on CDA (usually $<20\%$). One possible explanation is that such a simulation approach generates unstable rounding strategies and corresponding quantized networks at every step, making it much more challenging to identify a clear and convergent optimization direction than straightforward quantization-conditioned backdoors. Besides, the simulated rounding strategies are still different from EFRAP's final strategy, making the adaptive attack less robust against EFRAP. As the security research on backdoor vulnerabilities is an evolving game between attacks and defenses, we leave the study on more effective attacks to future work.

\section{More Visualization Results}
In this section, we provide more visualization results, including GradCAM \cite{selvaraju2017grad} and t-SNE \cite{van2008visualizing}.

\begin{figure*}[ht]
  \centering
  \begin{subfigure}[b]{\textwidth}
    \includegraphics[width=\textwidth]{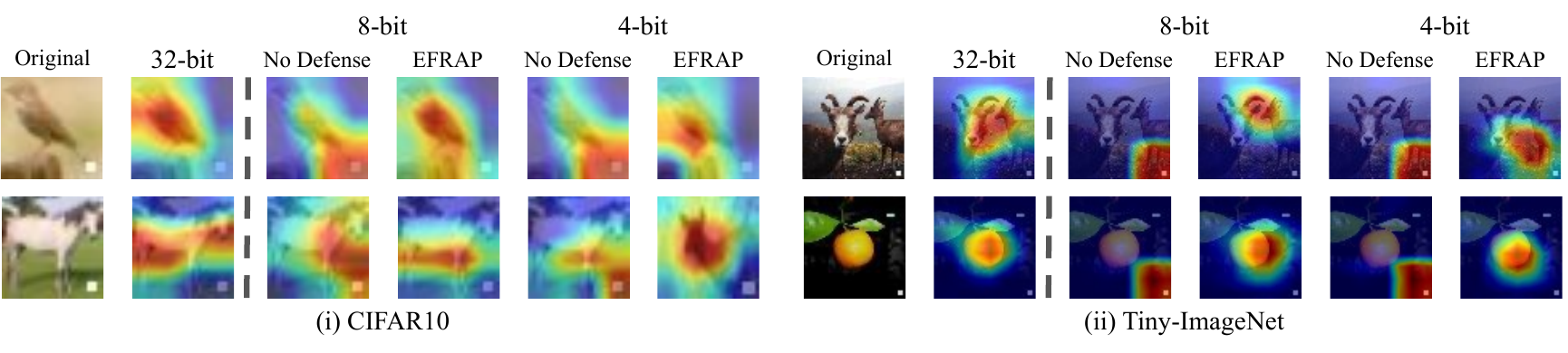}
    \caption{GradCAM results on CompArtifact \cite{tian2022stealthy}. It uses a 3$\times$3 small white patch as trigger for all datasets.}
    \label{fig:gradcam-1}
  \end{subfigure}
  \hfill
  \begin{subfigure}[b]{\textwidth}
    \includegraphics[width=\textwidth]{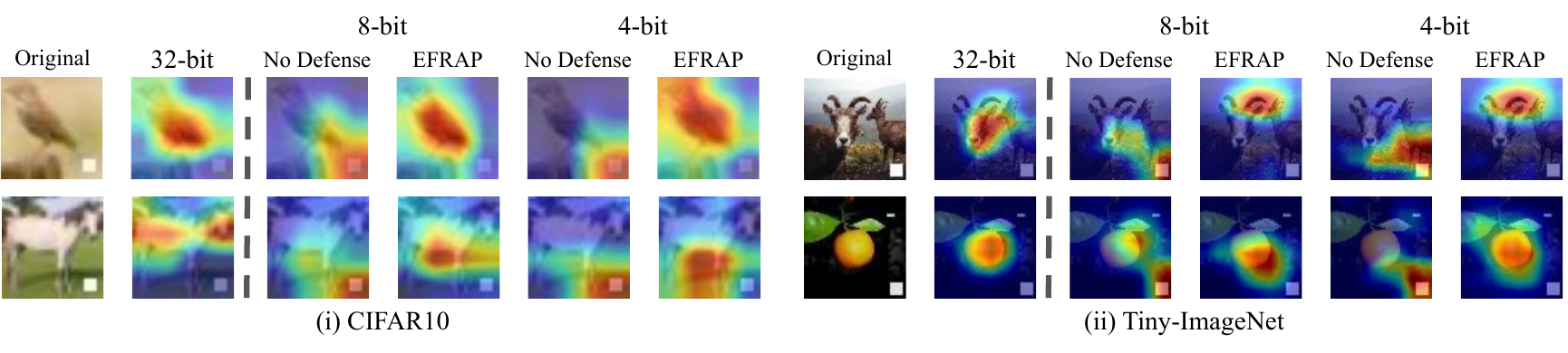}
    \caption{GradCAM results on Qu-ANTI-zation \cite{hong2021qu}. The trigger size is set to 4$\times$4 on CIFAR10 and 8$\times$8 on Tiny-ImageNet.}
  \end{subfigure}
  \hfill
  \begin{subfigure}[b]{\textwidth}
    \includegraphics[width=\textwidth]{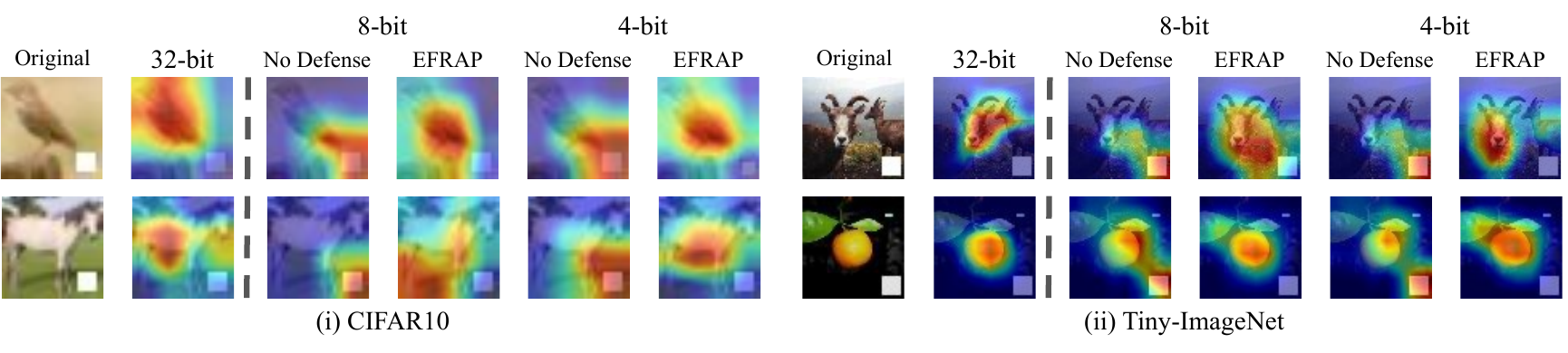}
    \caption{GradCAM results on PQBackdoor \cite{ma2021quantization,ma2023commercial}. The trigger size is set to 6$\times$6 on CIFAR10 and 12$\times$12 on Tiny-ImageNet.}
  \end{subfigure}
  \hfill
  \begin{subfigure}[b]{\textwidth}
    \includegraphics[width=\textwidth]{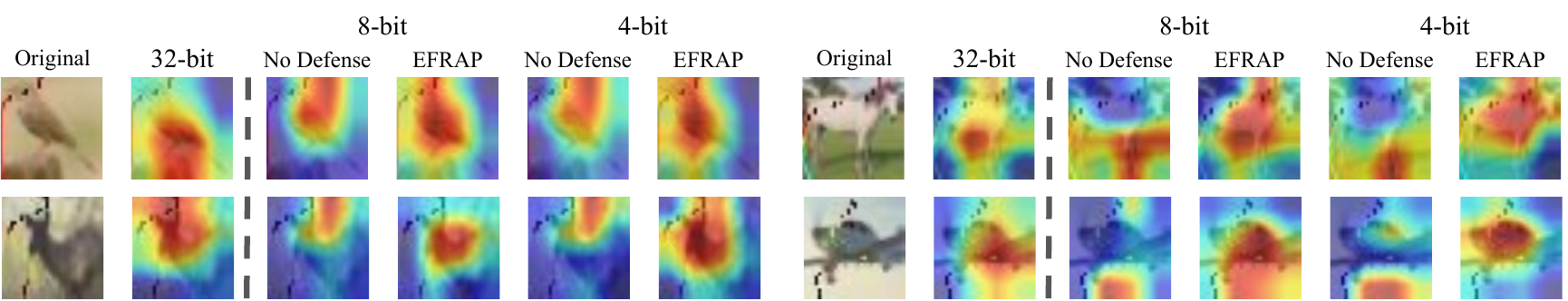}
    \caption{GradCAM results on input-aware dynamic trigger. We adopt PQBackdoor \cite{ma2021quantization,ma2023commercial} as the base attack.}
  \end{subfigure}
  \hfill
  \begin{subfigure}[b]{\textwidth}
    \includegraphics[width=\textwidth]{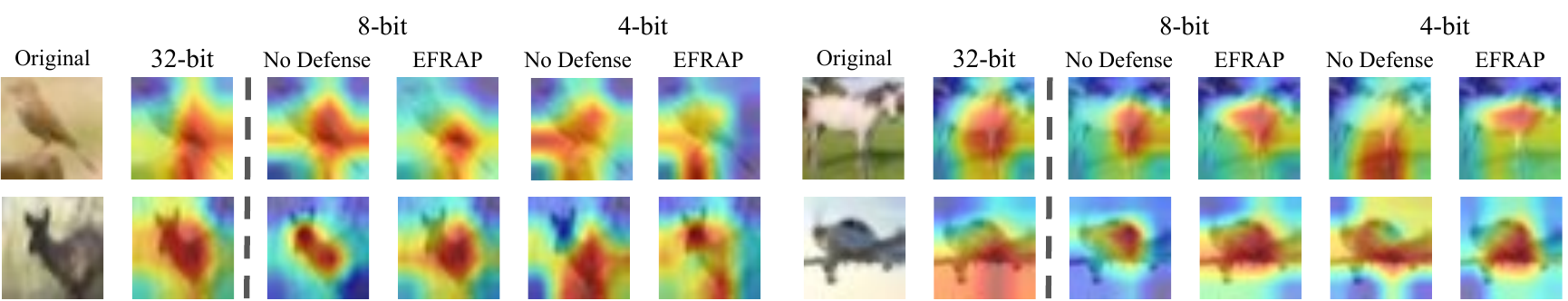}
    \caption{GradCAM results on warping-based trigger. We adopt PQBackdoor \cite{ma2021quantization,ma2023commercial} as the base attack.}
    \label{fig:gradcam-5}
  \end{subfigure}
  \caption{\textbf{More GradCAM Results.}}
  \label{fig:ablation}
\end{figure*}

\vspace{5pt}
\noindent\textbf{More GradCAM \cite{selvaraju2017grad} Results.} We provide more GradCAM results for each attack, including CompArtifact, Qu-ANTI-zation, and PQBackdoor, on CIFAR10 and Tiny-ImageNet, and PQBackdoor with advanced trigger (input-aware dynamic and warping-based) on CIFAR10, with 8-bit and 4-bit bandwidth, before and after defense. The results are shown in Figure \ref{fig:gradcam-1} to \ref{fig:gradcam-5}. The GradCAM results also demonstrate the effectiveness of EFRAP. After defense, the networks' activation focuses on the main object of the image, rather than the trigger area on the input. 

\begin{figure*}[ht]
  \centering
  \begin{subfigure}[b]{\textwidth}
    \includegraphics[width=\textwidth]{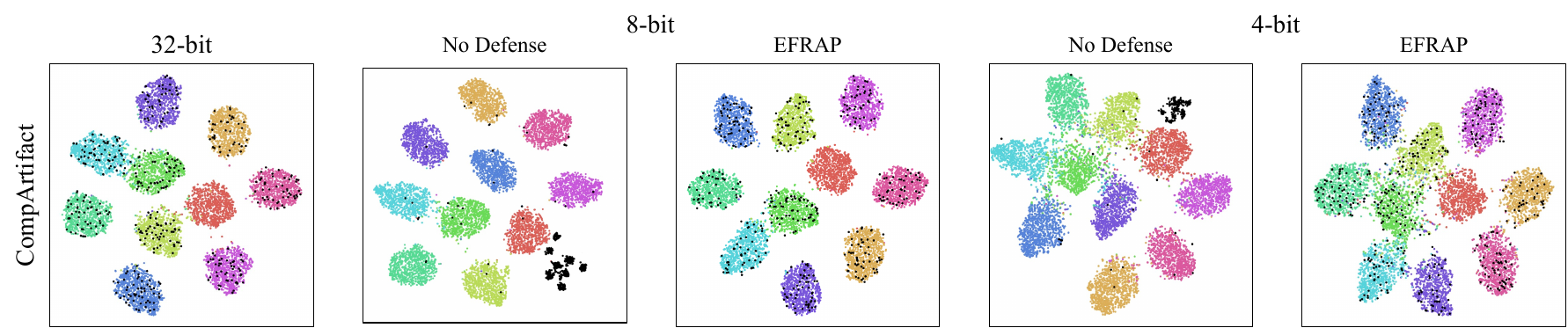}
    \caption{t-SNE results on CompArtifact \cite{tian2022stealthy}.}
    \label{fig:tsne-1}
  \end{subfigure}
  \hfill
  \begin{subfigure}[b]{\textwidth}
    \includegraphics[width=\textwidth]{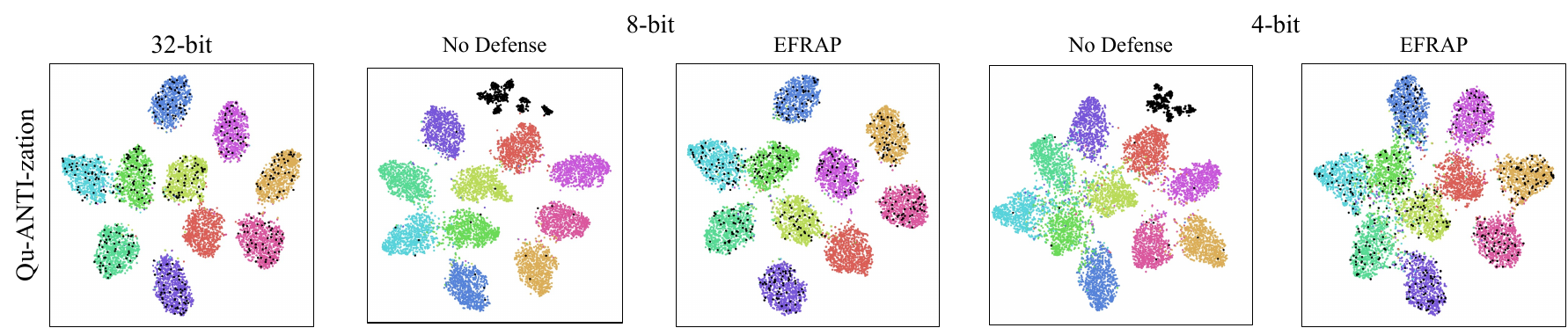}
    \caption{t-SNE results on Qu-ANTI-zation \cite{hong2021qu}.}
  \end{subfigure}
  \hfill
  \begin{subfigure}[b]{\textwidth}
    \includegraphics[width=\textwidth]{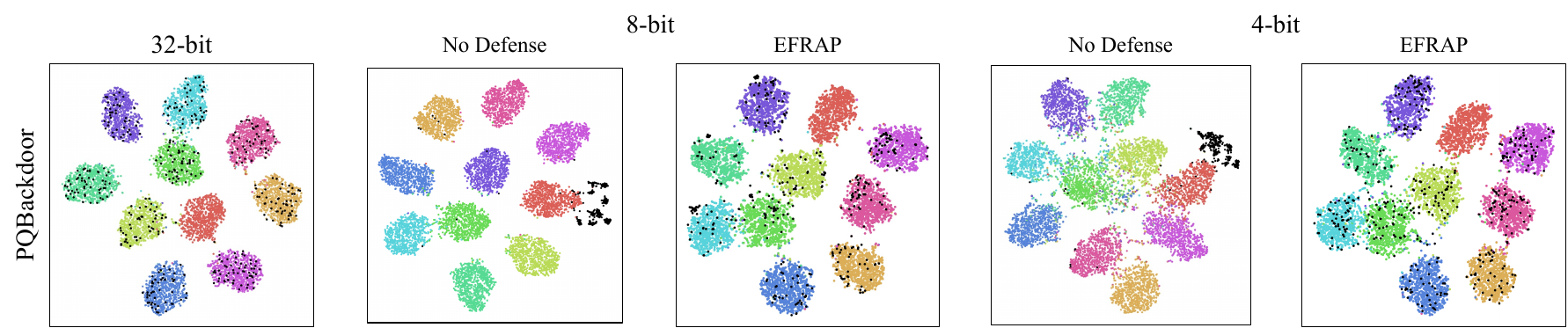}
    \caption{t-SNE results on PQBackdoor \cite{ma2021quantization,ma2023commercial}.}
    \label{fig:tsne-5}
  \end{subfigure}
  \caption{\textbf{More t-SNE Results.} }
\end{figure*}

\vspace{5pt}
\noindent\textbf{More t-SNE \cite{van2008visualizing} Results.} We provide more t-SNE results for each attack, including CompArtifact, Qu-ANTI-zation, and PQBackdoor, on CIFAR10, with 8-bit and 4-bit bandwidth, before and after defense.  The results are shown in Figure \ref{fig:tsne-1} to \ref{fig:tsne-5}. After EFRAP, the poisoned samples effectively disperse to their original category. This shows that EFRAP has successfully removed the backdoor effects in the model.

\section{Discussions}
\noindent\textbf{Ethical Statements.} The study of the security vulnerabilities of deep learning models has the potential to give rise to ethical concerns \cite{carlini2023poisoning,liu2022complex,walmer2022dual}. In this paper, for the first time, we propose a novel defense against the recently proposed quantization-conditioned backdoor attacks. We are confident that our method will strengthen the security of model quantization process, and safeguard the responsible deployment of deep learning models. We have carefully checked the CVPR 2024 Ethics Guidelines for Authors\footnote{https://cvpr.thecvf.com/Conferences/2024/EthicsGuidelines} and we are confident our research adheres to all mentioned ethical standards. We ensure that our methodologies and experiments do not harm individuals or organizations and comply with all relevant ethical guidelines and regulatory standards. Our defense mechanism, EFRAP, is intended solely for protecting DNNs against malicious tampering and is not designed for any unethical or harmful applications.

\begin{figure}[t]
    \centering
    \begin{minipage}{0.473\textwidth}
        \centering
        \includegraphics[width=0.8\textwidth]{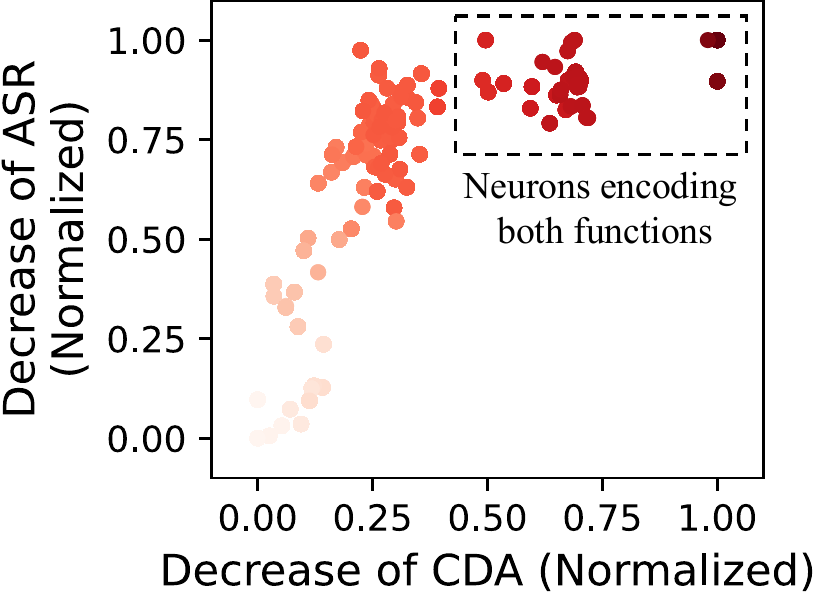}
        \caption{\textbf{Neuron Function Study.}}
        \label{fig:neuron}
    \end{minipage}\hfill
\end{figure}

\vspace{0.3em}
\noindent\textbf{Statistical Analysis on the Neuron's `Dual Encoding' Phenomenon.} We test the reduction of CDA/ASR after pruning each neuron with top 10\% error. As in Fig.~\ref{fig:neuron}, they mostly have high relations with ASR, while some of them are also key for CDA. This fact suggests that {{some neurons encode both backdoor and clean functions}}. This result aligns with Fine-Pruning \cite{liu2018fp}.

\vspace{0.3em}
\noindent\textbf{Other Implementations of Activation Preservation.} The primary objective of the activation preservation term in EFRAP is to compensate for benign accuracy after error-guided flipped rounding. Except for the activation MSE loss by \citet{nagel2020up}, many other alternative losses can be chosen for this purpose, $e.g.$, FlexRound \cite{lee2023flexround}, FIM-based Minimization \cite{li2021brecq}, Prediction Difference Metric \cite{liu2023pd}, or any other losses that can improve post-training quantization and is compatible for the 0-1 integer programming optimization. Currently, our activation preservation directly follows the work of \cite{nagel2020up}, and it is highly effective in compensating for clean accuracy drop caused by $\mathcal{L}_F$. As suggested by \cite{li2021brecq}, changing layer-wise activation preservation to block-wise can allow a more flexible optimization. We study a case on ResNet-18, PQBackdoor, and indeed {{a slight improvement}} (around $0.3  \%$ on CIFAR10) {{is observed}}. Interestingly, we have experimentally observed that these losses, although originally designed for mitigating accuracy loss during quantization, can mitigate the quantization-conditioned backdoors in some cases (but we did not do comprehensive experiments to verify this). It would be interesting to further discover these mechanisms in future works.

\vspace{0.3em}
\noindent\textbf{Defenses Results using Straight-Through Estimator (STE).} In our early trials, we have also considered applying existing defenses on quantized models via STE. However, as in Tab.~\ref{tab:quantized-results} and \ref{tab:quantized-results4}, {{the results are still discouraging}}. The most possible reasons are: (1) STE returns only coarse gradients, not perfectly accurate ones; (2) as the model is already trained with QAT (which already involved STE) by the attacker, the gradients of quantized and full-precision models are similar overall. Therefore the optimization directions are also similar in general, making a significant improvement less likely. 
\begin{table}[t]
\begin{minipage}[t]{0.5\textwidth}
    \begin{minipage}{1\textwidth}
        \centering
        \footnotesize
        \setlength{\tabcolsep}{4pt}
        \caption{\textbf{Results w/ STE on 8-bit PQ Backdoor.}}
        \scalebox{1}{
            \begin{tabular}{c c c}
            \toprule
                Defense  &  CDA / ASR \\
                \midrule
                FT + STE & 84.38 / 95.41 \\
                MCR + STE & 84.39 / 65.09 \\
                NAD + STE & 38.00 / {\color{white}0}5.86 \\
                I-BAU + STE & 82.05 / 19.58 \\
            \bottomrule
            \end{tabular}
        }
        \label{tab:quantized-results}
    \end{minipage}
\end{minipage}
\end{table}

\begin{table}[t]
\begin{minipage}[t]{0.5\textwidth}
    \begin{minipage}{1\textwidth}
        \centering
        \footnotesize
        \setlength{\tabcolsep}{4pt}
        \caption{ \textbf{Results w/ STE on 4-bit PQ Backdoor.}}
        \scalebox{1}{
            \begin{tabular}{c c c}
            \toprule
                Defense  &  CDA / ASR \\
                \midrule
                FT + STE & 82.95 / 93.12 \\
                MCR + STE & 82.16 / 40.29 \\
                NAD + STE & 39.67 / 11.17 \\
                I-BAU + STE & 76.15 / 20.80 \\
            \bottomrule
            \end{tabular}
        }
        \label{tab:quantized-results4}
    \end{minipage}
\end{minipage}
\end{table}

\vspace{0.3em}
\noindent\textbf{Limitations and Future Work.} Although we have comprehensively discussed the effectiveness of EFRAP under different settings, there are still some limitations. For example, EFRAP takes the place of standard quantization and aims to quantize the infected model without activating the backdoors. However, this also means that EFRAP cannot be used in conjunction with current state-of-the-art quantization techniques, probably limiting the effectiveness of EFRAP on keeping high clean data accuracy (especially on low-bit quantization or larger models). It would be interesting to discover post-quantization defenses that can operate after quantization in future works.

